\documentclass[aoas]{imsart}

\RequirePackage{amsthm,amsmath,amsfonts,amssymb}
\RequirePackage[authoryear]{natbib}
\RequirePackage[colorlinks,citecolor=blue,urlcolor=blue]{hyperref}

\usepackage{fix-cm}
\usepackage{graphicx}
\usepackage{enumerate}
\usepackage{mathabx}
\usepackage{bbm}
\usepackage{tikz}
\usepackage{array}
\usepackage[table]{xcolor}
\usepackage{relsize} 
\usepackage{booktabs}
\usetikzlibrary{shapes.geometric, arrows, positioning, arrows.meta, calc}

\tikzset{
   -Latex,auto,node distance =1 cm and 1 cm,semithick,
   el/.style = {inner sep=2pt, align=left, sloped}
}

\renewcommand*{\P}{\mathbbm{P}}
\newcommand*{\E}{\mathbbm{E}}

\renewcommand*{\vec}[1]{\boldsymbol{#1}}

\newcommand*{\set}[1]{\mathrm{#1}}

\newcommand*{\Ind}{\mathbbm{1}}

\newcommand*{\giv}{\;|\;}
\newcommand*{\mgiv}{\;\middle|\;}
\newcommand{\ind}{\perp\!\!\!\perp} 
 
\newcommand{\notinfect}{\tikz\draw[blue, line width=0.4pt] (0,0) circle (2.8pt);}
\newcommand{\infect}{\tikz\draw[red, fill=red] (0,0) rectangle ++(0.20,0.20);}

\startlocaldefs
\theoremstyle{plain}

\theoremstyle{definition}

\theoremstyle{plain}

\newtheorem{thm}{Theorem}
\newtheorem{corollary}{Corollary}
\theoremstyle{definition}
\newtheorem{asm}{Assumption}

\newtheorem{tcnd}[asm]{Simplifying Assumption}

\endlocaldefs

\begin{document}

\begin{frontmatter}
\title{A Counterfactual Framework for Estimating Infectious Disease Prevalence under Repeated Testing with Symptomatic and Contact-Tracing Components}
\runtitle{Repeated Testing with Symptomatic and Contact-tracing Components}

\begin{aug}

\author[A]{\fnms{Jeongjin} \snm{Lee}\ead[label=e0]{lee.10449@osu.edu}},
\author[A]{\fnms{Junke} \snm{Yang}\ead[label=e1]{yang.8009@osu.edu}},
\author[A]{\fnms{Grzegorz A.} \snm{Rempala}\ead[label=e2]{rempala.3@osu.edu}}
\and
\author[A,B]{\fnms{Patrick M.} \snm{Schnell}\ead[label=e3]{schnell.31@osu.edu}}

\address[A]{Division of Biostatistics, College of Public Health,
The Ohio State University\printead[presep={, }]{e0,e1,e2,e3}}
\address[B]{Department of Medical Epidemiology and Biostatistics, Karolinska Institutet}
\end{aug}

\begin{abstract}
This paper addresses the problem of estimating infectious disease prevalence under longitudinal testing programs that include scheduled, symptomatic, and contact-tracing testing.
Our study is motivated by data from The Ohio State University, where a mandatory once-per-week COVID-19 testing and isolation program was implemented during the Fall 2020 semester, supplemented by additional testing for symptomatic individuals and identified contacts.
In this setting, the probability of being tested depends on symptoms or contact-tracing status, creating a complex observation process.
We develop a counterfactual framework that links the observation process to a hypothetical process in which infection is prevented.
This formulation enables unbiased estimation of disease prevalence by modeling the testing process, possibly nonparametrically, without requiring explicit modeling of transmission dynamics, even though the testing and infection processes are jointly dependent.
\end{abstract}


\end{frontmatter}

\section{Introduction} 

Since the beginning of the COVID-19 pandemic in early 2020, a wide range of institutions implemented longitudinal testing and isolation programs to help monitor and mitigate transmission. 
These efforts were carried out in settings such as colleges and universities \citep{school21, college21, chang2021repeat}, workplace environments \citep{work22}, and professional sports organizations \citep{mba21}.
Although the primary objective of these regimens was to mitigate transmission by identifying and isolating infectious individuals, the resulting testing data were also used to estimate the prevalence and incidence of infection within the monitored population. 
Accurate prevalence estimation is essential for institutional risk assessment, such as evaluating outbreak potential or determining isolation capacity, and for informing public health policy decisions \citep{baker2022successful}.

A simple and widely used estimator of prevalence is the \textit{test-positive rate} (TPR), defined as the proportion of positive tests among all tests conducted on a given day. 
The TPR on a given day has often been interpreted as an estimate of prevalence on that same day \citep{kahanec2021impact}. 
When the tested sample is representative of the population and tests are perfectly sensitive and specific, the TPR provides an unbiased estimate of the true prevalence. 
However, as demonstrated by \citet{schnell2024overcoming}, the TPR exhibits a systematic upward bias under longitudinal testing with isolation, even when symptom-based and contact-tracing testing are absent.
This bias arises because the probability of being tested on a given day depends on the time since an individual’s last test, which is also correlated with the probability of being infectious. 
As a result, individuals tested on a given day are not generally representative of the monitored population, so prevalence estimation becomes a nonprobability sampling problem in which testing acts as a selection mechanism.

In the absence of symptomatic testing and contact tracing, \citet{schnell2024overcoming} proposed an unbiased Horvitz--Thompson (HT) estimator for prevalence, which uses inverse probability weighting to account for the testing mechanism.
The key idea behind the estimator is the analogy between how the testing process unfolds conditional on an individual not being infectious and how the process would unfold in a hypothetical world where nobody becomes infected.
This analogy was never formally justified.
Additionally, it was unclear how symptomatic testing and contact tracing could be addressed because of the already complex dependency between the longitudinal testing and infection processes.
Although the connection between survey sampling and causal inference is well known and has been discussed in the literature, see, for example, \citet{feinberg2018interlocking} and \citet{kohler2019nonprobability}, its application to longitudinal testing settings has been less explicit.
By using a counterfactual framework \citep{hernan2020causal}, we formalize the assumptions that justify this analogy in longitudinal settings.
We further use tools from causal inference such as directed acyclic graphs (DAGs) and single-world intervention graphs (SWIGs) to clarify how bias arises and to develop an estimator that handles additional complications such as symptomatic testing and contact tracing.

The remainder of this paper is organized as follows. 
Section~\ref{sec:framework} formally describes the joint disease and testing processes. 
Section~\ref{sec:schedule} introduces the counterfactual framework in causal inference and shows how it applies to the setting in \citet{schnell2024overcoming} without symptomatic testing and contact tracing.
Section~\ref{sec:preferential} applies these tools to develop a Horvitz-Thompson estimator for unbiased prevalence estimation with symptomatic testing and contact tracing. 
Section~\ref{sec:simulation} reports simulation results evaluating the finite-population performance of the proposed estimator under various testing designs, and Section~\ref{sec:realdata} applies the method to longitudinal COVID-19 testing data from The Ohio State University. 
Section~\ref{sec:discussion} concludes with methodological implications, limitations, and directions for future research.

\section{Formulation}
\label{sec:framework}

We describe the joint progression of disease and testing processes using a \textit{State-Based} framework, which provides a clear representation of population states and transitions at discrete time points. At any given time $t$, the population is partitioned into three mutually exclusive states: Well ($\set{W}(t)$), Infectious ($\set{I}(t)$), and Removed ($\set{R}(t)$).  
The indicator $W_i(t)$ is 1 if individual $i$ is in the Well state at time $t>0$, and 0 otherwise.
Indicators $I_i(t)$ and $R_i(t)$ are defined similarly.
This framework differs in key aspects from some common compartmental models, such as the SIR (Susceptible-Infectious-Removed) model \citep{kermack1927contribution}.
For instance, individuals in the Well state ($\set{W}(t)$) are not necessarily susceptible; they may be vaccinated or retain temporary immunity after recovering from a previous infection.
Similarly, individuals in the Removed state ($\set{R}(t)$) are not necessarily deceased or fully recovered but may simply be excluded from the actively monitored population for various reasons.
The total number of individuals in the Well state at time $t$ is given by $W_+(t) = \sum_{i=1}^{M} W_i(t),$ and similarly for the Infectious and Removed states, yielding $I_+(t)$ and $R_+(t)$, respectively. 
We denote the total population size as $M = W_+(t) + I_+(t) + R_+(t)$.

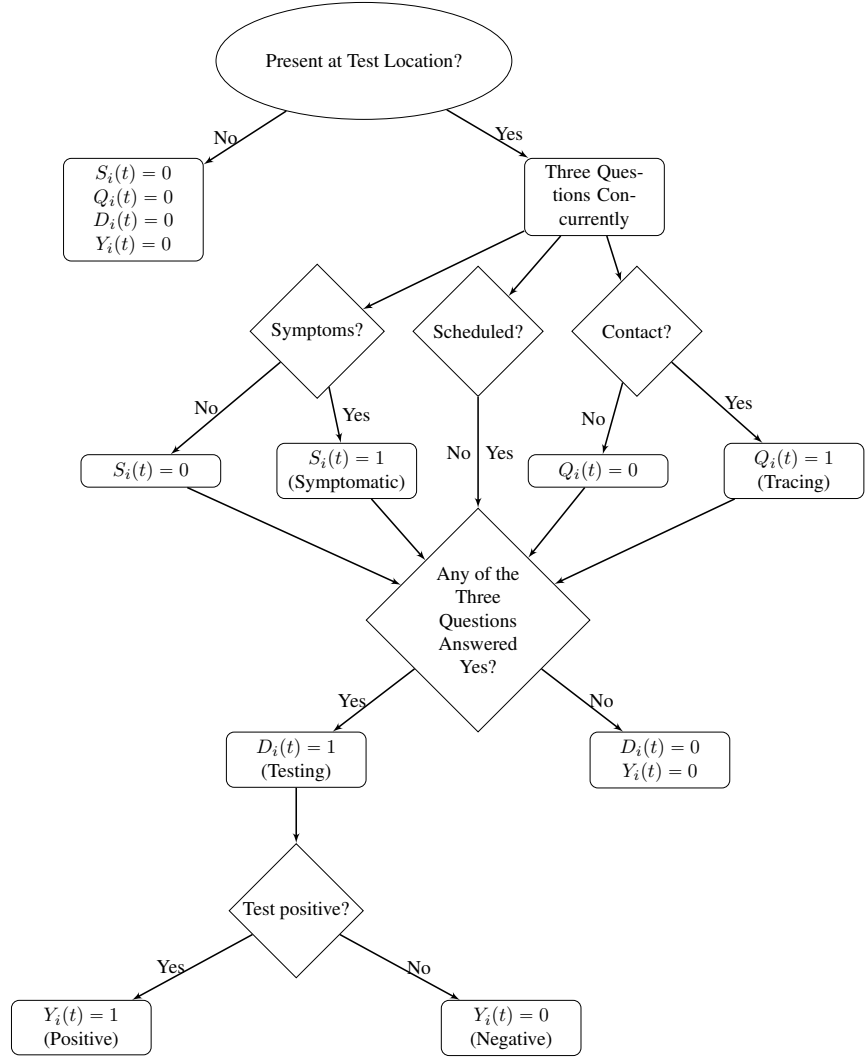
\begin{figure}
\small
\centering
\resizebox{0.8\textwidth}{!}{
\begin{tikzpicture}[
  node distance=1cm and 1cm, 
  every node/.style={align=center, font=\small},
  decision/.style={diamond, draw, text width=6em, text badly centered, node distance=0.5cm, inner sep=0pt},
  block/.style={rectangle, draw, text width=6.5em, text centered, rounded corners, minimum height=1em},
  line/.style={draw, -latex', thick},
  cloud/.style={draw, ellipse, node distance=1cm, minimum height=6em}
]

\node (start) [cloud] {Present at Test Location?}; 
\node (noLocation) [block, below left=of start] {$S_i(t) = 0$ \\$Q_i(t) = 0$ \\ $D_i(t) = 0$ \\ $Y_i(t) = 0$ };
\node (yesLocation) [block, below right=of start] {Three Questions Concurrently};

\node (sympt) [decision, below=of yesLocation, xshift=-5cm] {Symptoms?};
\node (sched) [decision, right=of sympt, node distance=4cm] {Scheduled?};
\node (trace) [decision, right=of sched, node distance=4cm] {Contact?};

\node (S0) [block, below=of sympt, xshift=-3cm] {$S_i(t) = 0$};
\node (S1) [block, right=of S0, node distance=4cm] {$S_i(t) = 1$ \\ (Symptomatic)};
\node (T0) [block, right=of S1, node distance=4cm, xshift=1cm] {$Q_i(t) = 0$};
\node (T1) [block, right=of T0, node distance=4cm] {$Q_i(t) = 1$ \\ (Tracing)};

\node (Yes_any) [decision, below =of sched, yshift=-1.5cm]{Any of the Three Questions Answered Yes?};
\node (YesTest) [block, below left= of Yes_any]{$D_i(t) = 1 $\\ (Testing)};
\node (NoTest) [block, below right= of Yes_any]{$D_i(t) = 0$ \\ $Y_i(t) = 0$};
\node (outcome) [decision, below=of YesTest, yshift=-0.5cm] {Test positive?};
\node (positive) [block, below left=of outcome, xshift=-1cm] {$Y_i(t) = 1$\\(Positive)};
\node (negative) [block, below right=of outcome, xshift=1cm] {$Y_i(t) = 0$\\(Negative)};

\path [line] (start) -- node[left] {No} (noLocation);
\path [line] (start) -- node[right] {Yes} (yesLocation);

\path [line] (yesLocation) -- (sched);
\path [line] (yesLocation) -- (sympt);
\path [line] (yesLocation) -- (trace);

\path [line] (sympt) -- node[left] {No} (S0);
\path [line] (sympt) -- node[right] {Yes} (S1);

\path [line] (trace) -- node[left] {No} (T0);
\path [line] (trace) -- node[right] {Yes} (T1);

\path [line] (S1) -- (Yes_any);
\path [line] (T1) -- (Yes_any);
\path [line] (S0) -- (Yes_any);
\path [line] (T0) -- (Yes_any);
\path [line] (sched) -- node[left] {No} node[right] {Yes} (Yes_any);

\path [line] (Yes_any) -- node[left] {Yes} (YesTest);
\path [line] (Yes_any) -- node[right] {No} (NoTest);

\path [line] (YesTest) -- (outcome);
\path [line] (outcome) -- node[left] {Yes} (positive);
\path [line] (outcome) -- node[right] {No} (negative);

\end{tikzpicture}
}
\caption{Flowchart for COVID-19 Testing}
\label{fig:flowchart}
\end{figure}

Transitions between states are governed by both infection and testing processes. 
Let $X_{il}$ denote the time of the $l$th exposure of individual $i$, and define $X_i=\min_l X_{il}$ as the first exposure time. 
This first exposure time determines the transition from the Well state to the Infectious state, so that individual $i$ moves from $\set{W}(X_i)$ to $\set{I}(X_i+1)$. 
If individual $i$ is never exposed, we define $X_i=\infty$, indicating that the individual remains in the Well state indefinitely.
The set of individuals tested at time $t$ is denoted by $\set{D}(t)$, where $\set{D}$ represents ``diagnostic" testing. 
This set includes scheduled testing, which refers to a systematic testing process that occurs regardless of symptoms or contact tracing.  
For individual $i$, let $Z_{ik}$ denote the time of the $k$th test.
Scheduled testing can be administratively assigned by a central scheduler or chosen by the individuals in the population themselves, provided they adhere to certain constraints.
For instance, scheduled testing may be based on a protocol that mandates a minimum of one test per week, even in the absence of any symptoms or known exposure through contact tracing. 
In addition to scheduled testing, the testing $\set{D}$ can be influenced by other factors such as symptoms or contact tracing.
Here we let the set $\set{S}(t)$ consist of individuals presenting symptoms at a testing location at time $t$.
The set $\set{Q}(t)$ comprises individuals identified through contact tracing at time $t$, based on evidence of recent exposure to an individual likely to be infectious. 
Membership in $\set{S}(t)$ and $\set{Q}(t)$ is identified through responses to specific questions administered at the test location.

The flowchart in Figure~\ref{fig:flowchart} provides a detailed illustration of the testing process. 
Among those tested, individuals who test positive are represented by the outcome $Y_i(t) = 1$.
For individuals not present at the test location, their testing-related variables are defined as follows: $S_i(t) = 0$, indicating no symptoms; $Q_i(t) = 0$, indicating no identification through contact tracing; $D_i(t) = 0$, indicating no testing; and $Y_i(t) = 0$, indicating no outcomes. 
These definitions ensure that the values of these variables are defined even for individuals who do not attend testing.
To improve readability, Table~\ref{tab:notation} summarizes the main notation used throughout the paper. 
It includes the state indicators, testing and symptom related variables, observed histories, and the key counterfactual outcome used in the identification argument. 
Throughout, \textit{counterfactual outcomes} (also called \textit{potential outcomes}) \citep{hernan2020causal} denote the hypothetical outcomes that would be observed under specified interventions.

\begin{table}[t]
\centering
\caption{Summary of main notation.}
\label{tab:notation}
\scriptsize
\begin{tabular}{p{0.25\textwidth} p{0.67\textwidth}}
\toprule
\textbf{Symbol} & \textbf{Meaning} \\
\midrule
$M$ & Population size. \\

$T$ & End of follow-up. \\

$W_i(t), I_i(t), R_i(t)$ & Indicators that individual $i$ is Well, Infectious, or Removed at time $t$. \\

$X_{il}$ & Time of the $l$th exposure of individual $i$. \\

$\vec{X}_i$ & Exposure history of individual $i$ over follow-up, defined by $\vec{X}_i = \{X_{i1},\dots,X_{iT}\}$. \\

$X_i(t)$ & Cumulative exposure indicator for individual $i$ by time $t$, defined by $X_i(t)=1$ if individual $i$ has been exposed by time $t$, and $X_i(t)=0$ otherwise. \\

$D_i(t)$ & Indicator that individual $i$ is tested at time $t$. \\

$Z_{ik}$ & Time of the $k$th test for individual $i$. \\

$\vec{Z}_i$ & Testing history of individual $i$ over follow-up, defined by $\vec{Z}_i=\{Z_{i1},\dots,Z_{iT}\}$. \\

$S_i(t)$ & Indicator that individual $i$ presents symptoms at time $t$. \\

$Q_i(t)$ & Indicator that individual $i$ is identified through contact tracing at time $t$. \\

$Y_i(t)$ & Test outcome indicator at time $t$, with $Y_i(t)=1$ for a positive result. \\

$\bar{X}_i(t), \bar{D}_i(t), \bar{S}_i(t), \bar{Q}_i(t), \bar{Y}_i(t)$ 
& Histories of exposure, testing, symptom presentation, contact tracing, and test outcomes for individual $i$ up to time $t$, respectively, where for example $\bar{X}_i(t)=\{X_i(1),\dots,X_i(t)\}$. \\

$D_i^{\bar{x}_{t-1}=0}(t)$ & Counterfactual testing indicator at time $t$ under no prior infection up to time $t-1$. \\
\bottomrule
\end{tabular}
\end{table}

\subsection{Assumptions}
This section introduces the assumptions underlying our model of infection and testing dynamics. The ``No undetected recoveries" assumption is essential. 
Other assumptions such as ``Perfect test sensitivity and specificity", ``No Clearance", and ``Identically distributed joint processes between individuals" are adopted purely for pedagogical reasons. 
An estimator that relaxes these simplifying assumptions will be introduced later, providing a framework more aligned with real-world conditions.

\begin{asm}[No undetected recoveries]
  \label{asm:no-undetected-recoveries}
  Individuals in $\set{I}$ cannot return to $\set{W}$ except by passing through $\set{R}$.
\end{asm}

\begin{tcnd}[Perfect test sensitivity and specificity]
  \label{asm:perfect-test}
  \begin{align}
    \P[Y_i(t) = 1 \giv D_i(t) = 1, I_i(t) = 1] &= 1, \quad \P[Y_i(t) = 1 \giv D_i(t) = 1, I_i(t) = 0] = 0.
  \end{align}
\end{tcnd}

\begin{tcnd}[No Clearance]
  \label{asm:no-clearance}
  Individuals who enter $\set{R}$ remain in $\set{R}$ forever.
\end{tcnd}

\begin{tcnd}[Identically distributed joint exposure and testing processes across individuals]
  \label{tcnd:id-between-individuals}
  For all $i \neq j$,
  \begin{equation}
    \P[\vec{x}_i, \vec{z}_i] = \P[\vec{x}_j, \vec{z}_j],
  \end{equation}
  where $\vec{x}_i$  denotes the exposure history of individual $i$ over follow-up and $\vec{z}_i$ denotes the corresponding testing history.
\end{tcnd}

Assumption~\ref{asm:no-undetected-recoveries} implies that all infections are eventually detected, which is plausible when testing intervals are short relative to the infectious period.
Assumption~\ref{asm:perfect-test} requires perfect test accuracy, ensuring infectious individuals always test positive and noninfectious individuals always test negative.
Together with Assumption~\ref{asm:no-undetected-recoveries}, Assumption~\ref{asm:no-clearance} ensures that individuals leaving $\set{W}$ do not return.
Finally, Assumption~\ref{tcnd:id-between-individuals} allows dependencies across individuals, as correlated testing or infection may occur within social or geographic clusters.

\section{A New Perspective on Scheduled Testing}
\label{sec:schedule}

In this section, we revisit the estimator of \cite{schnell2024overcoming} for prevalence under scheduled testing via a causal perspective.
This approach clarifies the justifications of certain assumptions and techniques, and will allow for extensions to symptomatic and contact tracing testing in the next section.

\subsection{Bias of Test-Positive Rate}
Estimating disease prevalence in a longitudinal testing context is subject to several biases, with a primary source of bias arising from the testing schedule itself, even in the absence of direct indicators of infectiousness, such as symptoms. 
A widely recognized and commonly used estimator in these contexts is the test-positive rate (TPR), which is defined as the proportion of positive tests among all tests conducted. 
However, as shown in \citet{schnell2024overcoming} and illustrated with a toy example in Table \ref{tab:weekly_testing}, the TPR can be a biased estimator of prevalence in the nonremoved population under periodic testing schemes. 
In the example shown in Table \ref{tab:weekly_testing}, individuals are tested randomly once per week, and removed if they test positive. 
While the test-positive rate remains one-third each day, the prevalence steadily decreases as infectious individuals are detected and subsequently isolated. 
This type of schedule testing creates a dependency between an individual’s likelihood of being tested and their probability of being infectious. 

\begin{table}[h!]
    \centering
    \tiny
    \renewcommand{\arraystretch}{1.5}
    \begin{tabular}{>{\centering}p{2cm}|*{5}{>{\centering\arraybackslash}p{1.5cm}}}
        \rowcolor{gray!40} \textbf{Person} & \textbf{Monday} & \textbf{Tuesday} & \textbf{Wednesday} & \textbf{Thursday} & \textbf{Friday} \\ \hline
        1 & \cellcolor{orange!30} \notinfect & \notinfect & \notinfect & \notinfect & \notinfect \\ \hline
        2 & \cellcolor{orange!30} \notinfect & \notinfect & \notinfect & \notinfect & \notinfect \\ \hline
        3 & \cellcolor{orange!30}\infect & & & & \\ \hline
        4 & \infect & \cellcolor{orange!30}\infect & & & \\ \hline
        5 & \notinfect & \cellcolor{orange!30}\notinfect & \notinfect & \notinfect & \notinfect \\ \hline
        6 & \notinfect & \cellcolor{orange!30}\notinfect & \notinfect & \notinfect & \notinfect \\ \hline
        7 & \infect & \infect & \cellcolor{orange!30} \infect \\ \hline
        8 & \notinfect & \notinfect & \cellcolor{orange!30}\notinfect & \notinfect & \notinfect \\ \hline
        9 & \notinfect & \notinfect & \cellcolor{orange!30}\notinfect & \notinfect & \notinfect \\ \hline
        10 & \infect & \infect & \infect & \cellcolor{orange!30}\infect \\ \hline
        11 & \notinfect & \notinfect & \notinfect & \cellcolor{orange!30}\notinfect & \notinfect \\ \hline
        12 & \notinfect & \notinfect & \notinfect & \cellcolor{orange!30} \notinfect & \notinfect \\ \hline
        13 & \notinfect & \notinfect & \notinfect & \notinfect & \cellcolor{orange!30} \notinfect \\ \hline
        14 & \notinfect & \notinfect & \notinfect & \notinfect & \cellcolor{orange!30} \notinfect \\ \hline
        15 & \infect & \infect & \infect & \infect & \cellcolor{orange!30} \infect \\ \hline
        \multicolumn{6}{c}{} \\ 
        \multicolumn{1}{c}{} & \multicolumn{3}{c}{\notinfect \quad Not Infectious \quad\quad \infect \quad Infectious} & \cellcolor{orange!30} Persons Tested  \\
    \end{tabular}
    \caption{Weekly schedule testing (Once per week testing) with Infectious status for 15 individuals. 
    Test-positive rate is one-third everyday, but prevalence decreases as infectious individuals are detected and removed, leading to bias.} 
    \label{tab:weekly_testing}
\end{table}

\citet{schnell2024overcoming} show that under perfect testing (Assumption~\ref{asm:perfect-test}) and identically distributed individuals (Assumption~\ref{tcnd:id-between-individuals}), for the TPR to serve as an unbiased estimator of prevalence, $\P\left[I(t)=1 \mid R(t)=0\right],$ it is necessary and sufficient that the infectiousness and testing probability of all nonremoved individuals at any time $t$ are independent.
This condition is known as the \textit{marginal independence of testing and infectiousness} (MITI), which is formally expressed  in our notation as: $D_i(t) \ind X_i(t) \mid R_i(t) = 0,$ where $X(t)$ denotes the exposure indicator at time $t$.
As illustrated above, MITI is violated by the once-per-week testing employed by OSU.

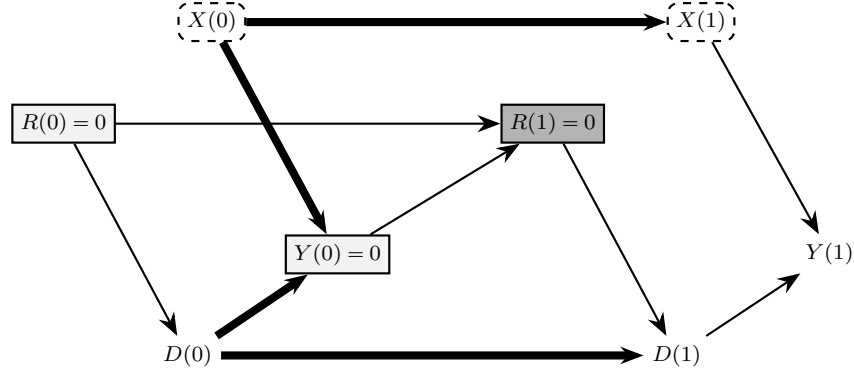
\begin{figure}[h!]
    \centering
    \resizebox{0.8\textwidth}{!}{
    \begin{tikzpicture}[
        node distance=1cm and 1.2cm,
        every node/.style={font=\scriptsize}, 
        condition/.style={draw, thick, font=\scriptsize, align=center}, 
        unobserved/.style={draw, thick, font=\scriptsize, align=center, dashed, rounded corners}, 
        edge/.style={->, thick, >={Stealth[length=3mm]}}
    ]
        \node[condition, fill=gray!10] (R0) at (0,0) {$R(0) = 0$};
        \node[unobserved, above right=0.8cm and 0.8cm of R0] (X0) {$X(0)$};
        \node[condition, below right=2.5cm and 0.5cm of X0, fill=gray!10] (Y0) {$Y(0)=0$};
        \node[below right=2.5cm and 0.5cm of R0] (D0) {$D(0)$};

        \node[condition, right=5cm of R0, fill=gray!60] (R1) {$R(1)=0$};
        \node[unobserved, above right=0.8cm and 0.8cm of R1] (X1) {$X(1)$};
        \node[below right=2.5cm and 0.8cm of X1] (Y1) {$Y(1)$};
        \node[below right=2.5cm and 0.5cm of R1] (D1) {$D(1)$};

        \draw[edge] (R0) -- (D0);
        \draw[edge] (R0) -- (R1);
        \draw[edge, line width=3pt] (D0) -- (Y0);
        \draw[edge] (Y0) -- (R1);
        \draw[edge] (R1) -- (D1);
        \draw[edge] (D1) -- (Y1);
        \draw[edge, line width=3pt] (X0) -- (Y0);
        \draw[edge] (X1) -- (Y1);
        \draw[edge, line width=3pt] (X0) -- (X1);
        \draw[edge, line width=3pt] (D0) -- (D1);
    \end{tikzpicture}
    }
    \caption{
    Directed Acyclic Graphs (DAGs) illustrating disease progression and testing at $t=0,1$. 
    The DAGs include exposure indicators $X(0)$ and $X(1)$. 
    Shaded nodes (solid = observed, dashed = unobserved) denote conditioning, including descendants of colliders. 
    Conditioning on $R(1)=0$ opens the path $D(1) \leftarrow D(0) \rightarrow Y(0) \leftarrow X(0) \rightarrow X(1)$, violating the MITI assumption and biasing the test-positive rate. 
    }
    \label{fig:Dag}
\end{figure}

The source of the test-positive rate bias can be understood formally using Directed Acyclic Graphs (DAGs). 
As illustrated in DAGs in Figure~\ref{fig:Dag}, the condition of MITI is not guaranteed to hold in longitudinal testing contexts. 
In particular, conditioning on $R(1)=0$ (a descendant of the collider $Y(0)$) opens a path between $D(1)$ and $X(1)$:
$D(1) \leftarrow D(0) \rightarrow Y(0) \leftarrow X(0) \rightarrow X(1).$


\subsection{Horvitz-Thompson Estimator}
In the previous subsection, we demonstrated how longitudinal testing regimens can introduce bias into the test-positive rate (TPR) as an estimator of prevalence. 
These biases often arise due to \textit{non-random sampling effects}, where individuals’ likelihood of being selected for testing is influenced by factors such as the time since their last test or recent test outcomes, rather than simple random testing.

To address this issue, we employ the Horvitz-Thompson (HT) estimator \citep{horvitz1952generalization}, which corrects for non-random sampling through inverse-probability weighting. 
Originally developed for survey sampling, the HT estimator provides an unbiased estimate of a population total by weighting each observed unit by the inverse of its inclusion probability. 
Let $\mathcal{P} = \{1, \dots, M\}$ denote a finite population with variable of interest $K_i$ for unit $i$, and let $s \subset \mathcal{P}$ be a sample with inclusion probabilities $\pi_i = \P(i \in s) > 0$. 
The estimator for the total $K_+ = \sum_{i=1}^M K_i$ is $\hat{K}_{+}^{\text{HT}} = \sum_{i \in s} \frac{K_i}{\pi_i}.$
When inclusion probabilities are correctly specified, 
the HT estimator is unbiased, $\E(\hat{K}_{+}^{\text{HT}}) = K_+$, even if $K_i$ are random, provided that $\pi_i$ are defined conditional on the realized $\vec{K}$.

The survey sampling perspective is also relevant in longitudinal settings, where inclusion probabilities may vary over time.
Related work has examined sampling schemes for longitudinal binary response data in which selection depends on observed outcomes, outcome history, or related auxiliary variables \citep{schildcrout2008outcome, schildcrout2011outcome, schildcrout2012outcome}.
For example, \citet{schildcrout2008outcome} proposed a retrospective outcome dependent sampling design for longitudinal binary response data, in which subjects are selectively sampled on the basis of their observed response histories to permit efficient estimation of regression parameters for costly time varying exposures.
A key distinction in our setting is that testing is not simply an observation mechanism. 
Rather, it can directly affect the target process itself, namely the subsequent prevalence, because positive test results can lead to removal or isolation and thus change future prevalence.
Our contribution is to develop an identification and estimation framework for prevalence under repeated testing, where the probability of inclusion in the tested sample may depend on testing history.
From this perspective, the HT estimator provides a natural approach to estimating $W_+(t)=\sum_{i=1}^M W_i(t)$.

To formalize this, let $\set{s}(t) = \{ i : D_i(t) = 1 \}$ denote the set of individuals tested at time $t$.
Under perfect testing, $W_i(t)$ is only observable for individuals included in $\set{s}(t)$, as $W_i(t)$ is inferred from their observed test outcome $1 - Y_i(t)$.
Since the population Well status $\vec{W}(t) = (W_1(t),\ldots,W_M(t))$ can be random, the \textit{time-specific inclusion probability} for each individual $i$ is the conditional probability of being tested, given $\vec{W}(t)$:
\[
\pi_i(t) = \P[i \in \set{s}(t) \mid \vec{W}(t)] = \P[D_i(t)=1 \mid \vec{W}(t)].
\]
When these probabilities are correctly specified, the HT estimator, 
\begin{equation}
\hat{W}_{+}^{\text{HT}} 
= \sum_{i \in \set{s}(t)} \frac{W_i(t)}{\pi_i(t)} = \sum_{i \in \set{s}(t)} \frac{1 - Y_i(t)}{\pi_i(t)} 
= \sum_{i} \frac{\Ind(i \in \set{s}(t)) \cdot (1 - Y_i(t))}{\pi_i(t)} 
= \sum_{i} \frac{D_i(t)(1 - Y_i(t))}{\pi_i(t)},
\end{equation}
is unbiased for $W_+(t)$ conditional on $\vec{W}(t)$ (see Theorem~\ref{thm:s-unbiased-estimator}).
Moreover, conditioning on $\vec{W}(t)$ can be further refined. Since $W_i(t)$ is a binary variable, we have
\begin{equation}
\frac{W_i(t)} {\pi_i(t)} = \frac{W_i(t)} { \P[D_i(t)=1 \mid \vec{W}(t)]}  = \frac{W_i(t)} {\P[D_i(t)=1 \mid W_i(t)=1, \vec{W}_{-i}(t)]},
\label{eq:pi_reduction}
\end{equation}
where $\vec{W}_{-i}(t)$ denotes the vector of Well states for all individuals, excluding individual $i$, at time $t$.
The conditional probability $\P[D_i(t)=1 \mid W_i(t)=1, \vec{W}_{-i}(t)]$ will be used in the key assumptions introduced below.

We begin by outlining two key assumptions upon which the HT estimator is based and subsequently demonstrate its unbiasedness under these assumptions.

\begin{asm}[Independence of testing from others' states (ITOS)]
  \label{asm:s-itos}
  \begin{equation*}
    \P\left[D_{i}(t) = 1 \mgiv W_i(t) = 1, \vec{W}_{-i}(t) \right] = \P\left[D_{i}(t) = 1 \mgiv W_i(t) = 1 \right].
  \end{equation*}
\end{asm}
ITOS holds if and only if $D_i(t) \ind \vec{W}_{-i}(t) \mid W_i(t) = 1$, where $\vec{W}_{-i}(t)$ denotes the vector of Well states for all individuals, excluding individual $i$, at time $t$.
This independence is particularly evident in the absence of symptomatic testing or contact tracing, as scheduled testing depends solely on the individual's own state and is unaffected by the states of others in the population. 
Relative to the corresponding assumption in \citet{schnell2024overcoming}, the ITOS assumption here is imposed only for individuals with $W_i(t)=1$, rather than for both $W_i(t)=1$ and $W_i(t)=0$. 
This is sufficient for identification because, by Equation~\eqref{eq:pi_reduction}, only the testing probability among Well individuals enters the HT estimator.

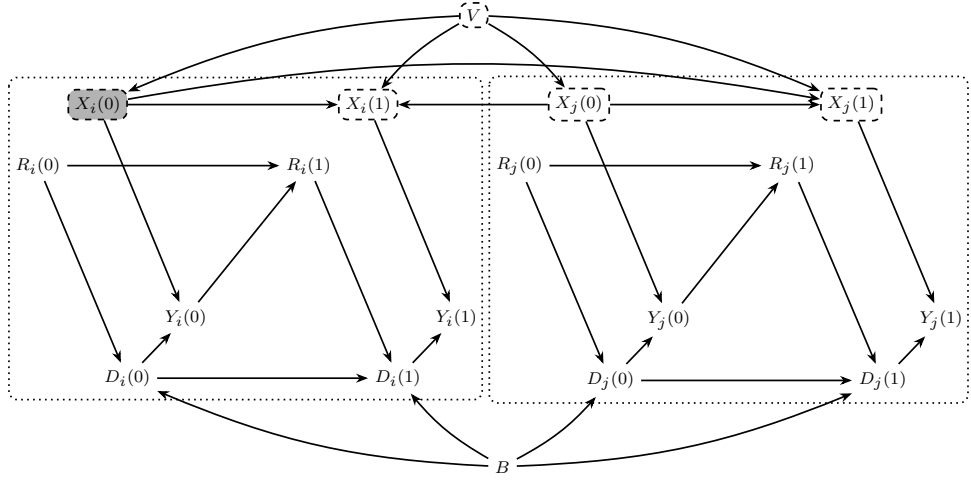
\begin{figure}[h]
    \centering
    \resizebox{0.9\textwidth}{!}{
    \begin{tikzpicture}[
        node distance=0.6cm and 1.2cm,
        every node/.style={font=\scriptsize}, 
        observed/.style={draw, thick, font=\scriptsize, align=center}, 
        unobserved/.style={draw, thick, font=\scriptsize, align=center, dashed, rounded corners}, 
        condition/.style={draw, thick, font=\scriptsize, align=center}, 
        dottedbox/.style={draw, thick, dotted, rounded corners},
        edge/.style={->, thick, >={Stealth[length=2mm]}}
    ]

        \node (R0i) at (0,0) {$R_i(0)$};
        \node[below right=2cm and 1.5cm of R0i] (Y0i) {$Y_i(0)$};
        \node[below right=3cm and 0.5cm of R0i] (D0i) {$D_i(0)$};
        \node[unobserved, above left=3cm and 0.5cm of Y0i, fill=gray!60] (X0i) {$X_i(0)$};

        \node[right=3.5cm of R0i] (R1i) {$R_i(1)$};
        \node[below right=2cm and 1.5cm of R1i] (Y1i) {$Y_i(1)$};
        \node[below right=3cm and 0.5cm of R1i] (D1i) {$D_i(1)$};
        \node[unobserved, condition, above left=3cm and 0.5cm of Y1i] (X1i) {$X_i(1)$};

        \node[below right=1cm and 1cm of D1i] (U) {$B$};
        \node[unobserved, above right =1cm and 1cm of X1i] (V) {$V$};

        \node[right=2.5cm of R1i] (R0j) {$R_j(0)$};
        \node[below right=2cm and 1.5cm of R0j] (Y0j) {$Y_j(0)$};
        \node[below right=3cm and 0.5cm of R0j] (D0j) {$D_j(0)$};
        \node[unobserved, above left=3cm and 0.5cm of Y0j] (X0j) {$X_j(0)$};

        \node[right=3.5cm of R0j] (R1j) {$R_j(1)$};
        \node[below right=2cm and 1.5cm of R1j] (Y1j) {$Y_j(1)$};
        \node[below right=3cm and 0.5cm of R1j] (D1j) {$D_j(1)$};
        \node[unobserved, above left=3cm and 0.5cm of Y1j] (X1j) {$X_j(1)$};

        \draw[edge] (X0i) -- (X1i);
        \draw[edge] (X0i) -- (Y0i);
        \draw[edge] (X1i) -- (Y1i);
        \draw[edge] (R0i) -- (D0i);
        \draw[edge] (R0i) -- (R1i);
        \draw[edge] (D0i) -- (Y0i);
        \draw[edge] (Y0i) -- (R1i);
        \draw[edge] (R1i) -- (D1i);
        \draw[edge] (D1i) -- (Y1i);
        \draw[edge] (D0i) -- (D1i);

        \draw[edge] (X0j) -- (X1j);
        \draw[edge] (X0j) -- (Y0j);
        \draw[edge] (X1j) -- (Y1j);
        \draw[edge] (R0j) -- (D0j);
        \draw[edge] (R0j) -- (R1j);
        \draw[edge] (D0j) -- (Y0j);
        \draw[edge] (Y0j) -- (R1j);
        \draw[edge] (R1j) -- (D1j);
        \draw[edge] (D1j) -- (Y1j);
        \draw[edge] (D0j) -- (D1j);

        \draw[edge, bend left=10] (U) to (D0i);
        \draw[edge, bend left=10] (U) to (D1i);
        \draw[edge, bend right=10] (U) to (D0j);
        \draw[edge, bend right=10] (U) to (D1j);    

        \draw[edge, bend right=10] (V) to (X0i);
        \draw[edge, bend right=10] (V) to (X1i);
        \draw[edge, bend left=10] (V) to (X0j);
        \draw[edge, bend left=10] (V) to (X1j); 

        \draw[edge, bend left=10] (X0i) to (X1j);
        \draw[edge] (X0j) to (X1i);
        
        \draw[dottedbox] ($(R0i.north west)+(0,1.2)$) rectangle ($(D1i.south east)+(0.9, -0.1)$);
        \draw[dottedbox] ($(R0j.north west)+(0,1.2)$) rectangle ($(D1j.south east)+(0.9, -0.1)$);
    \end{tikzpicture}
    }
    \caption{Directed Acyclic Graphs (DAGs) illustrating disease progression and testing dynamics over $t=0$ and $t=1$ for individuals $i$ and $j$, incorporating unobserved variables $V$ and observed shared circumstance $B$, which influence Well states and testing behaviors, respectively. Shaded nodes show conditioned variables, dashed borders indicate unobserved variables, and dotted boxes visually separate the nodes associated with individuals $i$ and $j$. The testing $D_i(1)$ for individual $i$ remains conditionally independent of $X_j(1)$ when conditioned on $X_i(0) = 0$, demonstrating the principle of ITOS.}
    \label{fig:Dag.itos}
\end{figure}

Consider a scenario with two individuals, $i$ and $j$, as illustrated in Figure~\ref{fig:Dag.itos}.  
At time $t = 1$, the ITOS assumption is expressed as:
$\P\left[D_i(1) = 1 \mid W_i(1) = 1, W_j(1)\right] = \P\left[D_i(1) = 1 \mid W_i(1) = 1\right],$
which follows from the conditional independence, $D_i(1) \ind W_j(1) \mid W_i(1) = 1.$
This conditional independence can be equivalently written as
$D_i(1) \ind X_j(1) \mid X_i(0) = 0,$
since the Well status $W_i(1) = 1$ implies $X_i \geq 1$, which in turn implies $X_i(0) = 0$.  
Here, $X_i$ denotes the exposure time of individual $i$, and $X_i(t)$ is the exposure indicator.

Additionally, in Figure~\ref{fig:Dag.itos}, we introduce a new unobserved variable $V$ and an observed shared circumstance $B$, which influence all exposure time $X$ and all testings $D$, respectively. 
The variable $V$ represents unmeasured factors that affect the health states of individuals. 
Examples of $V$ include exposures to infectious diseases such as COVID-19 or the common cold, environmental conditions like air pollution or extreme weather, shared living or working environments that increase the likelihood of transmission of infectious agents, and genetic predispositions that influence individual susceptibility to illnesses. 
The variable $B$ captures observed contextual factors influencing testing behaviors. 
Examples of $B$ include shared circumstances such as students opting for COVID-19 testing after class due to the proximity of the testing site, institutional policies assigning specific testing days for groups, or social influences where peers encourage collective testing behavior.
Figure~\ref{fig:Dag.itos} shows that the testing indicator $D_i(1)$ for individual $i$ remains conditionally independent of $X_j(1)$ upon conditioning on $X_i(0)$.
Paths through $B$ encounter unconditioned colliders (e.g., $Y_j(0)$ and $Y_j(1)$) on the $j$ side and are thus blocked.
Paths through $V$ must pass through $X_i(0)$, which is directly conditioned on and not a collider, thereby blocking those paths.

\begin{asm}[Positivity]
  \label{asm:s-positivity}
  For any $t$, $\P[D_{i}(t) = 1 \giv W_i(t) = 1] > 0$.
\end{asm}
If violated, this assumption biases prevalence estimates, as some groups are systematically untested, leading to infinite weights and invalid estimation.

Theorem~\ref{thm:s-unbiased-estimator} provides an unbiased HT estimator for $W_+(t)$ using inverse testing probability weights. 
Rather than computing prevalence directly as $I_+(t) / [M - R_+(t)]$, we first estimate the total $W_+(t) = \sum_i W_i(t)$ and then use the known quantities $M$ and $R_+(t)$ to derive the corresponding prevalence estimate.

\begin{thm}[Unbiased estimator of prevalence]
    \label{thm:s-unbiased-estimator}
  Assume perfect test sensitivity and specificity (Assumption~\ref{asm:perfect-test}), no clearance (Assumption~\ref{asm:no-clearance}), independence of testing from others' states (ITOS, Assumption~\ref{asm:s-itos}), and positivity (Assumption~\ref{asm:s-positivity}).
  Let $\omega_i(t) = 1 / \P[D_{i}(t) = 1 \giv W_i(t) = 1]$. Then,
  \small{
  \begin{equation}
    \label{eq:s-estimator-expectation}
    \begin{aligned}
      \E\left[
        \sum_i \omega_i(t) D_i(t) \{1 - Y_i(t)\} 
        \mgiv \vec{W}(t)
      \right]
      = W_+(t).
    \end{aligned}
  \end{equation}
  }
\end{thm}

A full proof of Theorem~\ref{thm:s-unbiased-estimator} is provided in the appendix.
Briefly, under perfect sensitivity and specificity (Assumption~\ref{asm:perfect-test}), $\P\left[D_i(t) = 1, Y_i(t) = 0 \mid \vec{W}(t) \right]$ is equivalent to $\P \left[D_i(t) = 1, W_i(t) = 1 \mid \vec{W}(t) \right]$.
ITOS (Assumption~\ref{asm:s-itos}) and the fact that $W_i$ is binary allow conditioning on $W_i(t) = 1$ instead of $\vec{W}(t)$, and positivity (Assumption~\ref{asm:s-positivity}) ensures that $\omega_i(t) \P\left[D_i(t) = 1 \mid W_i(t) = 1\right] = 1$ for all $i$. We obtain $W_+(t) = \sum_i W_i(t)$, yielding an unbiased estimator.

Interestingly, the HT estimator can be interpreted as an Inverse Probability of Treatment Weighting (IPTW) estimator, with the ``treatment" defined as ``testing".
By weighting inversely to the testing probability given the Well state, the IPTW approach constructs a pseudo-population where testing is independent of infection, thereby blocking all backdoor paths between $D_i(t)$ and $N_i^{d(t)=1}(t)$, where $N_i(t)=D_i(t)\{1-Y_i(t)\}$ and $N_i^{d(t)=1}(t)$ denotes the counterfactual negative test outcome under universal testing ($d(t)=1$). 
As shown in the Appendix (Proposition~1), the IPTW estimator coincides with the HT estimator.

\subsection{Computing Testing Probabilities from a Counterfactual Scenario}

Accurately estimating $\P\left[D_i(t) = 1 \mid W_i(t) = 1\right]$ for all $i$ is essential for constructing an unbiased estimator. 
However, observing these probabilities directly is generally infeasible, even with a comprehensive understanding of the testing protocol, since $W_i(t)$ is only partially observed.
We will show that, under certain conditions, $\P\left[D_i(t)=1 \mid W_i(t)=1\right]$ equals $\P\left[D_i^{\bar{x}_{t-1}=0}(t)=1\right]$, the testing probability for individual $i$ in a counterfactual world in which all prior exposure indicators are set to zero.
The following assumptions (consistency and exchangeability) establish the connection between the counterfactual and factual worlds.

\begin{asm}[Consistency]
  \label{asm:s-consist}
  If $X_i = x$, then $D_i^{\text{obs}}(t) = D_i^{x}(t)$.
\end{asm}

Assumption~\ref{asm:s-consist} states that if the exposure time is $X_i = x$, then the observed testing indicator at time $t$, denoted $D_i^{\text{obs}}(t)$, is equal to the counterfactual value $D_i^{x}(t)$ under an intervention that sets the exposure time to $x$. 
This assumption formally links the observed and counterfactual testing indicators when the exposure time is set to its observed value.
Furthermore, if the exposure indicator history up to time $t$ is given by $\bar{X}_i(t-1) = \bar{x}_{t-1}$, where $\bar{X}_i(t-1) = (X_i(0), X_i(1), \ldots, X_i(t-1))$ and $X_i(s) = \Ind[X_i = s]$ for $s \leq t-1$, then the consistency assumption implies that $D_i^{\text{obs}}(t) = D_i^{\bar{x}_{t-1}}(t).$

\begin{asm}[Exchangeability]
  \label{asm:s-exch}
  \begin{equation*}
      D_i^{\bar{x}_{t-1} = 0}(t) \ind \{X_i(0), X_i^{x_0=0}(1), \ldots, X_i^{x_{t-2}=0}(t-1) \}
  \end{equation*}
\end{asm}

Assumption~\ref{asm:s-exch} states that the counterfactual testing indicator at time $t$ under an intervention that sets exposure to zero up to time $t-1$, denoted $D_i^{\bar{x}_{t-1} = 0}(t)$, is independent of counterfactual exposure indicators up to time $t-1$.
This is stronger than a simpler condition such as $D_i^{x_{t-1}=0}(t) \ind X_i(t-1)$, because it excludes dependence on any earlier exposure, not only on the most recent one, which is necessary when testing decisions may be affected by past exposure history.

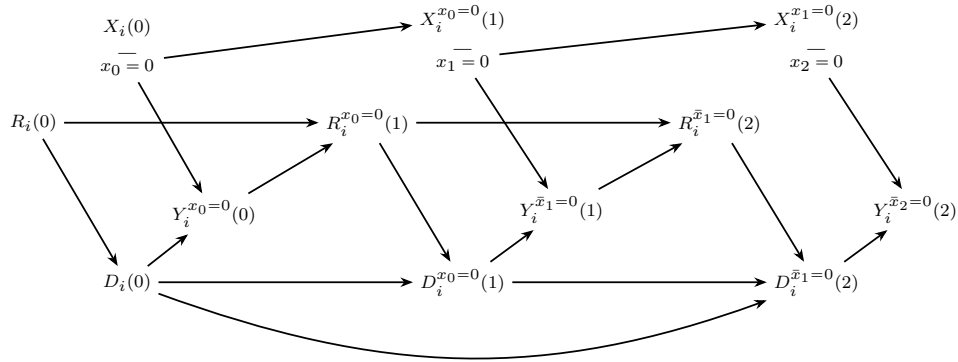
\begin{figure}[h!]
    \centering
    \resizebox{0.9\textwidth}{!}{
    \begin{tikzpicture}[
        node distance=0.6cm and 0.8cm,
        every node/.style={font=\scriptsize}, 
        condition/.style={draw, thick, font=\scriptsize, align=center}, 
        unobserved/.style={draw, thick, font=\scriptsize, align=center, dashed, rounded corners}, 
        dottedbox/.style={draw, thick, dotted, rounded corners},
        edge/.style={->, thick, >={Stealth[length=2mm]}}
    ]
        \node (Ri0) at (0,0) {$R_i(0)$};
        \node[below right=2cm and 0.5cm of Ri0] (Di0) {$D_i(0)$};
        \node[above right=0.5cm and 0.1cm of Di0] (Yi0) {$Y_i^{x_0=0}(0)$};

        \node[right=4cm of Ri0] (Ri1) {$R_i^{x_0 = 0}(1)$};
        \node[right=4cm of Di0] (Di1) {$D_i^{x_0 = 0}(1)$};
        \node[above right=0.5cm and 0cm of Di1] (Yi1) {$Y_i^{\bar{x}_1=0}(1)$};

        \node[right=4cm of Ri1] (Ri2) {$R_i^{\bar{x}_{1} = 0}(2)$};
        \node[right=4cm of Di1] (Di2) {$D_i^{\bar{x}_{1} = 0}(2)$};
        \node[above right=0.5cm and 0.0cm of Di2] (Yi2) {$Y_i^{\bar{x}_2=0}(2)$};

        \node[above=3.5cm of Di0] (Xi0) {$X_i(0)$};
        \node[below=0 of Xi0] (xi0) {$\genfrac{}{}{0pt}{}{\rotatebox{90}{$\mid$}}{\mathlarger{x_0=0}}$};
        
        \node[above=3.5cm of Di1] (Xi1) {$X_i^{x_0=0}(1)$};
        \node[below=0 of Xi1] (xi1) {$\genfrac{}{}{0pt}{}{\rotatebox{90}{$\mid$}}{\mathlarger{x_1=0}}$};
        
        \node[above=3.5cm of Di2] (Xi2) {$X_i^{x_1=0}(2)$};
        \node[below=0 of Xi2] (xi2) {$\genfrac{}{}{0pt}{}{\rotatebox{90}{$\mid$}}{\mathlarger{x_2=0}}$};
        
        \draw[edge] (Ri0) -- (Di0);
        \draw[edge] (Ri0) -- (Ri1);
        \draw[edge] (Ri1) -- (Ri2);
        \draw[edge] (Di0) -- (Yi0);
        \draw[edge] (Ri1) -- (Di1);
        \draw[edge] (Di1) -- (Yi1);
        \draw[edge] (Di0) -- (Di1);
        \draw[edge] (Di1) -- (Di2);
        \draw[edge] (Ri2) -- (Di2);
        \draw[edge] (Di2) -- (Yi2);       
        \draw[edge] (xi0) -- (Yi0);        
        \draw[edge] (xi1) -- (Yi1);        
        \draw[edge] (xi2) -- (Yi2);
        \draw[edge] (xi0) -- (Xi1);
        \draw[edge] (xi1) -- (Xi2);
        \draw[edge] (Yi0) -- (Ri1);
        \draw[edge] (Yi1) -- (Ri2);

        \draw[edge] (Di0) to[out=340, in=200] (Di2);
        
    \end{tikzpicture}
    }
    \caption{Single World Intervention Graph (SWIG) illustrating disease progression and testing dynamics from $t=0$ to $t=2$ for individual $i$.
    The graph includes an intervention that sets exposure indicators $x_0 = x_1 = x_2 = 0$.
    Under this intervention, the exchangeability $D_i^{\bar{x}_1 = 0}(2) \ind \{X_i(0), X_i^{x_0 = 0}(1)\}$ holds, which represents a simple case of Assumption~\ref{asm:s-exch}.}
    \label{fig:SWIG.sched.1}
\end{figure}

The \textit{Single World Intervention Graph (SWIG)} framework \citep{hernan2020causal} is a graphical approach in causal inference that incorporates counterfactual outcomes.
To construct a SWIG (Figure~\ref{fig:SWIG.sched.1}) from a DAG (Figure~\ref{fig:Dag.itos}), the exposure indicator (e.g., $X_i(0)$) is split into its fixed value under the intervention (e.g., $x_0=0$) and a separate node representing its natural value, $X_i(0)$.
Any edges originally directed out of the exposure variable in the DAG are moved to be directed out of the intervened exposure $x_0 = 0$.
Any variables that are descendants of the intervention node $x_0=0$ are labeled as counterfactual outcomes (e.g., $Y_i^{x_0=0}(0)$, $R_i^{x_0=0}(0)$, $D_i^{x_0=0}(1)$).
The same construction applies for interventions at later times, such as $x_1=0$ and $x_2=0$.
The SWIG (Figure~\ref{fig:SWIG.sched.1}) depicts an intervention that sets the exposure indicators to $x_0 = x_1 = x_2 = 0$, i.e., no exposure occurs through time $2$.
The absence of any open (i.e., unblocked) path between $D_i^{\bar{x}_1 = 0}(2)$ and the counterfactual exposure variables $\{X_i(0), X_i^{x_0 = 0}(1)\}$ in the SWIG implies the exchangeability, $D_i^{\bar{x}_1 = 0}(2) \ind \{X_i(0), X_i^{x_0 = 0}(1)\},$ as stated in Assumption~\ref{asm:s-exch}.

Theorem~\ref{thm:s-testing probability} further establishes that the probability of testing, given that an individual is in the Well state at time $t$ (i.e., $W_i(t) = 1$), is equal to the counterfactual probability of testing under a hypothetical scenario where exposure does not occur.

\begin{thm}[Estimation of testing probability]
  \label{thm:s-testing probability}
  Assume perfect test sensitivity and specificity (Assumption~\ref{asm:perfect-test}), consistency (Assumption~\ref{asm:s-consist}), and exchangeability (Assumption~\ref{asm:s-exch}). Then $\P[D_{i}(t) = 1 \giv W_i(t) = 1] =  \P [D_i^{\bar{x}_{t-1} = 0}(t) = 1].$ 
\end{thm}

Theorem~\ref{thm:s-testing probability} addresses the challenge posed by the unobservability of $W_i(t)$ in real-world data by providing an equivalence. 
Instead of directly estimating $\P[D_i(t) = 1 \mid W_i(t) = 1]$, which relies on the unobservable state $W_i(t)$, we can estimate $\P[D_i^{\bar{x}_{t-1} = 0}(t) = 1]$, the counterfactual testing probability for individual $i$ under an intervention that fixes all prior exposure indicators to zero. 
This probability can be computed by simulating the testing mechanism in a hypothetical scenario where every individual is intervened on to prevent exposure at all previous time points. 
If the testing mechanism is fully known, such as in scenarios where it is under total administrative control and all tests are explicitly documented, $\P[D_i^{\bar{x}_{t-1} = 0}(t) = 1]$ can be directly derived from the protocol. 
However, if the testing mechanism is not fully known, estimating these probabilities requires using observed data.

\subsection{Estimating Counterfactual Testing Probabilities from Data}
When the testing mechanism is not fully known, $\P[D_i^{\bar{x}_{t-1} = 0}(t) = 1]$ cannot be directly derived from the protocol. Therefore, it becomes necessary to estimate testing probabilities using observable factual data. 
In this section, we establish the connection between the counterfactual and factual worlds, providing a framework to infer testing probabilities even in the absence of complete knowledge about the testing mechanism.

We define $K_i(t) = \max\{k : Z_{ik} < t\},$
so that $Z_{i,K_i(u+1)+1}$ denotes the time of the next test at or after $u+1$.
Then, $\P[D_i^{\bar{x}_{t-1} = 0}(t) = 1]$ can be decomposed as follows:
{\small
\begin{equation}
\begin{aligned}
\P\left[D_i^{\bar{x}_{t-1} = 0}(t) = 1\right] = \sum_{u=0}^{t-1} 
\left( 
  \P\left[Z_{i, K_i(u+1)+1}^{\bar{x}_{t-1} = 0} = t \mid D_i^{\bar{x}_{u-1} = 0}(u) = 1\right] 
  \times \P\left[D_i^{\bar{x}_{u-1} = 0}(u) = 1\right] 
\right),
\end{aligned}
\end{equation}
}
where we adopt the convention that $D_i^{\bar{x}_{-1} = 0}(0) = 1$ for all $i$. 
The probability $\P[D_i^{\bar{x}_{u-1} = 0}(u) = 1]$ is identifiable since, by the inductive hypothesis, all counterfactual testing probabilities prior to time $t$ (that is, for all $u<t$) are already identified.
The remaining task is to calculate $\P[Z_{i,K_i(u+1)+1}^{\bar{x}_{t-1} = 0} = t \mid D_i^{\bar{x}_{u-1} = 0}(u) = 1]$. 
Define the term $\P[Z_{i,K_i(u+1)+1}^{\bar{x}_{t-1} = 0} = \tau \mid Z_{i,K_i(u+1)+1}^{\bar{x}_{t-1} = 0} \geq \tau, D_i^{\bar{x}_{u-1} = 0}(u) = 1]$ as the counterfactual conditional hazard function for the next test, $h^{\bar{x}_{t-1} = 0}(\tau)$, under an intervention that fixes the exposure history up to time $t-1$ to zero.
The probability $\P[Z_{i,K_i(u+1)+1}^{\bar{x}_{t-1} = 0} = t \mid D_i^{\bar{x}_{u-1} = 0}(u) = 1]$ can be expressed as a product involving $h^{\bar{x}_{t-1} = 0}(\tau)$:
\begin{equation}
\begin{aligned}
\P&\left[Z_{i,K_i(u+1)+1}^{\bar{x}_{t-1} = 0} = t \mid D_i^{\bar{x}_{u-1} = 0}(u) = 1\right] = h^{\bar{x}_{t-1} = 0}(t) \times \prod_{\tau=u+1}^{t-1} 
\left( 1 -  h^{\bar{x}_{\tau-1} = 0}(\tau) \right). 
\end{aligned}
\end{equation}
The counterfactual conditional hazard can be equivalently defined as:
\begin{equation}
h^{\bar{x}_{\tau-1} = 0}(\tau) = \P[Z_{i,K_i(u+1)+1}^{\bar{x}_{\tau-1} = 0} = \tau \mid Z_{i,K_i(u+1)+1}^{\bar{x}_{\tau-1} = 0} \geq \tau, D_i^{\bar{x}_{u-1} = 0}(u) = 1, Y_i^{\bar{x}_u = 0}(u)=0],
\end{equation}
which holds because $Y_i^{\bar{x}_u = 0}(u) = 0$ under the assumption of perfect testing (Assumption~\ref{asm:perfect-test}).

The next step is to establish a link between the counterfactual conditional hazard function $h^{\bar{x}_{\tau-1} = 0}(\tau)$ and the factual conditional hazard function,
\begin{equation}
h^{\text{obs}}(\tau) = \P[Z_{i,K_i(u+1)+1} = \tau \mid Z_{i,K_i(u+1)+1} \geq \tau, D_i(u) = 1, Y_i(u) = 0],    
\end{equation}
observed in real-world data.
Theorem~\ref{thm:s-hazard-link} establishes the equivalence between counterfactual and factual conditional hazard functions, allowing the counterfactual conditional hazard function to be identified using observed (factual) data.

\begin{thm}[Equivalence of Counterfactual and Factual Conditional Hazard Functions]
\label{thm:s-hazard-link}
Under perfect test sensitivity and specificity (Assumption~\ref{asm:perfect-test}), consistency (Assumption~\ref{asm:s-consist}), and exchangeability (Assumption~\ref{asm:s-exch}), the counterfactual conditional hazard function $h^{\bar{x}_{\tau-1} = 0}(\tau)$ is equal to the corresponding factual conditional hazard function $h^{\text{obs}}(\tau)$ for any $u \in \{0, \ldots, t-1\}$ and $\tau \in \{u+1, \ldots, t\}$:
\small{
\begin{equation}
h^{\bar{x}_{\tau-1} = 0}(\tau) = h^{\text{obs}}(\tau),   
\end{equation}
}
where
\small{
\begin{equation}
\begin{aligned}
h^{\bar{x}_{\tau-1} = 0}(\tau) 
= \P\Big[Z_{i,K_i(u+1)+1}^{\bar{x}_{\tau-1} = 0} = \tau \mid Z_{i,K_i(u+1)+1}^{\bar{x}_{\tau-1} = 0} \geq \tau, D_i^{\bar{x}_{u-1} = 0}(u) = 1, Y_i^{\bar{x}_u = 0}(u) = 0 \Big],
\end{aligned}
\end{equation}
}
and
\small{
\begin{equation}
\begin{aligned}
h^{\text{obs}}(\tau) 
= \P\Big[ Z_{i,K_i(u+1)+1} = \tau \mid Z_{i,K_i(u+1)+1} \geq \tau, D_i(u) = 1, Y_i(u) = 0 \Big].
\end{aligned}
\end{equation}
}
\end{thm}

\begin{corollary}
  \label{cor:test-probs}
  Assume perfect test sensitivity and specificity (Assumption~\ref{asm:perfect-test}), consistency (Assumption~\ref{asm:s-consist}), and exchangeability (Assumption~\ref{asm:s-exch}). Then, the counterfactual probability of being tested at time $t$ under an intervention that sets all prior exposure indicators to zero can be expressed as:
  \small{
  \begin{equation}
  \begin{aligned}
  \P\left[D_i^{\bar{x}_{t-1} = 0}(t) = 1\right] 
  = \sum_{u = 0}^{t-1} 
  \left(
  \P\left[ Z_{i, K_i(u+1)+1} = t \mid D_i(u) = 1, Y_i(u) = 0 \right]
  \times \P\left[D_i^{\bar{x}_{u-1} = 0}(u) = 1\right]
  \right).
  \end{aligned}
  \end{equation}
  }
  Under the stated assumptions, each term on the right-hand side is identifiable from the observed data. 
\end{corollary}

Theorem~\ref{thm:s-hazard-link} (proof in Appendix) follows from iterative consistency and exchangeability, as justified by the SWIG in Figure~\ref{fig:SWIG.sched.1}.  
Corollary~\ref{cor:test-probs} then expresses the counterfactual testing probability in terms of observed data, using Theorem~\ref{thm:s-hazard-link}.
Specifically, $\P[Z_{i,K_i(u+1)+1}=t \mid D_i(u)=1, Y_i(u)=0]$ is estimable from observed variables, and $\P[D_i^{\bar{x}_{u-1}=0}(u)=1]$ can be obtained recursively from earlier observed data.

In this section, we extended the core framework introduced by \citet{schnell2024overcoming} by incorporating Directed Acyclic Graphs (DAGs) and the counterfactual outcomes framework. 
DAGs provide a formal representation of the dependencies among key processes, including infection dynamics, testing indicator, and testing outcome, allowing for clearer identification of sources of bias. 
The counterfactual outcomes framework offers a rigorous foundation for linking factual and counterfactual outcomes, thereby enabling more precise formulation and evaluation of the assumptions required for identification and unbiased estimation. 
The methods developed here provide the foundation for extending the framework to symptomatic and contact-tracing testing in the next section.

\section{Scheduled Testing with Symptom Representation and Contact Tracing}
\label{sec:preferential}

\begin{figure}[h]
    \centering
    \resizebox{0.9\textwidth}{!}{
    \begin{tikzpicture}[
        node distance=0.6cm and 1.2cm,
        every node/.style={font=\scriptsize}, 
        observed/.style={draw, thick, font=\scriptsize, align=center}, 
        unobserved/.style={draw, thick, font=\scriptsize, align=center, dashed, rounded corners}, 
        dottedbox/.style={draw, thick, dotted, rounded corners},
        edge/.style={->, thick, >={Stealth[length=2mm]}}
    ]

        \node (R0i) at (0,0) {$R_i(0)$};
        \node[below right=2cm and 1.5cm of R0i] (Y0i) {$Y_i(0)$};
        \node[above =0.5cm of D0i] (S0i) {$S_i(0)$};
        \node[below right=3cm and 0.5cm of R0i] (D0i) {$D_i(0)$};
        \node[unobserved, above left=3cm and 0.5cm of Y0i] (X0i) {$X_i(0)$};

        \node[right=3.5cm of R0i] (R1i) {$R_i(1)$};
        \node[below right=2cm and 1.5cm of R1i] (Y1i) {$Y_i(1)$};
        \node[below right=3cm and 0.5cm of R1i] (D1i) {$D_i(1)$};
        \node[unobserved, above left=3cm and 0.5cm of Y1i] (X1i) {$X_i(1)$};
        \node[above =0.5cm of D1i] (S1i) {$S_i(1)$};

        \node[below right=1.5cm and 1cm of D1i] (U) {$B$};
        \node[unobserved, above right =1cm and 1cm of X1i] (V) {$V$};

        \node[right=2.7cm of R1i] (R0j) {$R_j(0)$};
        \node[below right=2cm and 1.5cm of R0j] (Y0j) {$Y_j(0)$};
        \node[below right=3cm and 0.5cm of R0j] (D0j) {$D_j(0)$};
        \node[above =0.5cm of D0j] (S0j) {$S_j(0)$};
        \node[unobserved, above left=3cm and 0.5cm of Y0j] (X0j) {$X_j(0)$};

        \node[right=3.5cm of R0j] (R1j) {$R_j(1)$};
        \node[below right=2cm and 1.5cm of R1j] (Y1j) {$Y_j(1)$};
        \node[below right=3cm and 0.5cm of R1j] (D1j) {$D_j(1)$};
        \node[unobserved, above left=3cm and 0.5cm of Y1j] (X1j) {$X_j(1)$};
        \node[above =0.5cm of D1j] (S1j) {$S_j(1)$};

        \node[right =0.25cm of D1i] (Q1i) {$Q_i(1)$};
        \node[right =0.25cm of D1j] (Q1j) {$Q_j(1)$};

        \draw[edge] (X0i) -- (X1i);
        \draw[edge] (X0i) -- (Y0i);
        \draw[edge] (X1i) -- (Y1i);
        \draw[edge] (R0i) -- (D0i);
        \draw[edge] (R0i) -- (R1i);
        \draw[edge] (D0i) -- (Y0i);
        \draw[edge] (Y0i) -- (R1i);
        \draw[edge] (R1i) -- (D1i);
        \draw[edge] (D1i) -- (Y1i);
        \draw[edge] (D0i) -- (D1i);
        \draw[edge] (X1i) -- (S1i);
        \draw[edge] (S1i) -- (D1i);
        \draw[edge] (X0i) -- (S0i);
        \draw[edge] (S0i) -- (D0i);

        \draw[edge] (X0j) -- (X1j);
        \draw[edge] (X0j) -- (Y0j);
        \draw[edge] (X1j) -- (Y1j);
        \draw[edge] (R0j) -- (D0j);
        \draw[edge] (R0j) -- (R1j);
        \draw[edge] (D0j) -- (Y0j);
        \draw[edge] (Y0j) -- (R1j);
        \draw[edge] (R1j) -- (D1j);
        \draw[edge] (D1j) -- (Y1j);
        \draw[edge] (D0j) -- (D1j);
        \draw[edge] (X1j) -- (S1j);
        \draw[edge] (S1j) -- (D1j);
        \draw[edge] (X0j) -- (S0j);
        \draw[edge] (S0j) -- (D0j);

        \draw[edge] (Q1i) -- (D1i);
        \draw[edge] (Q1j) -- (D1j);
        \draw[edge, bend left=5] (Y0j) to (Q1i);
        \draw[edge, bend right=30] (Y0i) to (Q1j);

        \draw[edge, bend left=10] (U) to (D0i);
        \draw[edge, bend left=10] (U) to (D1i);
        \draw[edge, bend right=10] (U) to (D0j);
        \draw[edge, bend right=10] (U) to (D1j);    

        \draw[edge, bend right=10] (V) to (X0i);
        \draw[edge, bend right=10] (V) to (X1i);
        \draw[edge, bend left=10] (V) to (X0j);
        \draw[edge, bend left=10] (V) to (X1j);

        \draw[edge, bend left=10] (X0i) to (X1j);
        \draw[edge] (X0j) to (X1i);

        \draw[dottedbox] ($(R0i.north west)+(0,1.2)$) rectangle ($(Q1i.south east)+(-0.1, -0.1)$);
        \draw[dottedbox] ($(R0j.north west)+(0,1.2)$) rectangle ($(Q1j.south east)+(-0.1, -0.1)$);
    \end{tikzpicture}
    }
    \caption{Directed Acyclic Graphs (DAGs) illustrating disease progression and testing dynamics over $t=0$ and $t=1$ for individuals $i$ and $j$, incorporating variables $S_i(1)$, $S_j(1)$ (symptoms at $t=1$) and $Q_i(1)$, $Q_j(1)$ (contact tracing at $t=1$). Dashed borders represent unobserved variables, and dotted boxes visually separate the nodes associated with individuals $i$ and $j$.}
    \label{fig:Dag.symp}
\end{figure}
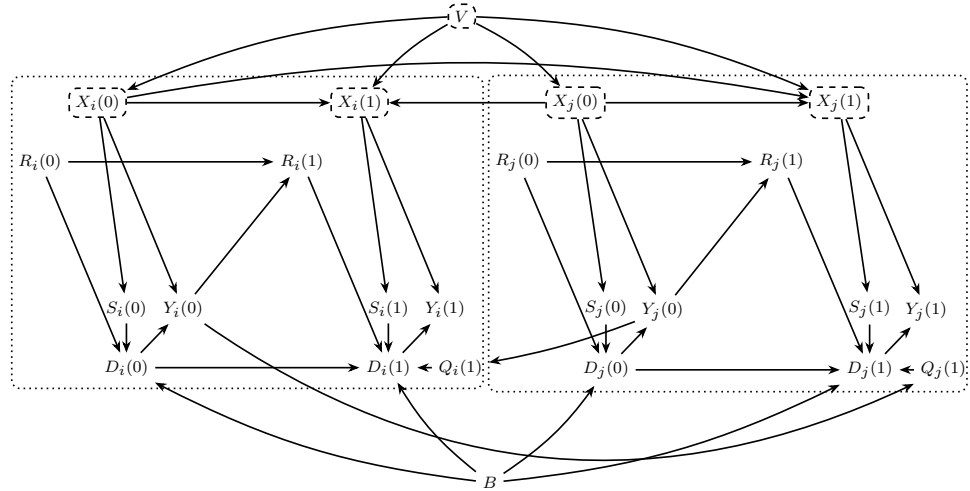

This section examines the entire timeline depicted in Figure~\ref{fig:flowchart}, which highlights key decision points where individuals present at the testing location are asked three questions.
Symptomatic individuals are more likely to seek testing and more likely to be infectious compared to the general population, so the naive Test-Positive Rate (TPR) is inflated relative to the true prevalence.  
Contact tracing introduces a similar distortion: individuals identified through recent exposure are more likely to be tested and thus become overrepresented among those tested.
As shown in Figure~\ref{fig:Dag.symp}, the \textit{marginal independence of testing and infectiousness} (MITI), $D_i(t) \ind X_i(t) \mid R_i(t) = 0$, is violated.  
MITI is violated due to the causal pathway depicted in Figure~\ref{fig:Dag.symp}:  
$X_i(t) \rightarrow S_i(t) \rightarrow D_i(t),$
where the exposure indicator $X_i(t)$ influences the likelihood of testing $D_i(t)$ through symptom representation $S_i(t)$.  
Since $X_i(t)$ determines $I_i(t)$ given that $R_i(t) = 0$, this violation undermines the validity of TPR as an unbiased estimator of prevalence even under simple random testing.

In Figures~\ref{fig:flowchart} and~\ref{fig:Dag.symp}, $S_i(t)$ and $Q_i(t)$ represent the presence of symptoms and contact tracing, respectively, both of which serve as causal factors influencing testing ($D_i(t)$). 
Below, we outline the formal criteria governing $S_i(t)$ and $Q_i(t)$:
\begin{enumerate}
    \item \textit{Exclusion Restriction}: There are no direct effects from $X_i(t)$ to $D_i(t)$ or from $Y_j(t-1)$ to $D_i(t)$. 
    The variable $X_i(t)$ influences $D_i(t)$ only through $S_i(t)$, and similarly, the variable $Y_j(t-1)$ influences $D_i(t)$ only through $Q_i(t)$.

    \item \textit{Observability}: Both $S_i(t)$ and $Q_i(t)$ are known for all individuals $i$, which can be ensured by either of the following mechanisms:
    \begin{enumerate}
        \item $S_i(t)$ (or $Q_i(t)$) is directly observed for all individuals $i$ (e.g., through mandatory reporting of symptoms or contact tracing information).
        \item When $D_i(t) = 1$, it implies that $S_i(t)$ (or $Q_i(t)$) is observed, and when $S_i(t) = 1$ (or $Q_i(t) = 1$), it implies that $D_i(t) = 1$.
    \end{enumerate}
\end{enumerate}
These requirements ensure that both $S_i(t)$ and $Q_i(t)$ are accurately defined and fully observed.
By convention, we set $S_i(0) = Q_i(0) = 0$, indicating that no individual is symptomatic or contact-traced at baseline. 
For each individual $i$, let $\bar{S}_i(t) = (S_i(0), \ldots, S_i(t))$ and $\bar{Q}_i(t) = (Q_i(0), \ldots, Q_i(t)),$ denote the histories of symptom indicators and contact-tracing eligibility up to time $t$, respectively, where each element takes values in $\{0,1\}$. 
The corresponding matrices $\vec{S}(t), \ \vec{Q}(t) \in \{0,1\}^{m \times (t+1)}$ collect these individual histories across all $m$ individuals, with each row representing $\bar{S}_i(t)$ or $\bar{Q}_i(t)$.

We now extend the framework to allow for non-perfect testing with symptom history, contact tracing history, and shared circumstance, modeling misclassification through random sensitivity and specificity \citep{schnell2024overcoming}.

\begin{asm}[Simple random sensitivity and specificity]
  \label{asm:simple-sens-spec}
  Positive test results are indicated by $Y_i(t) = D_i(t) \{F_i(t)(1-W_i(t)) + (1-G_i(t))W_i(t) \}$ with $F_i(t)$, $G_i(t)$ Bernoulli random variables with success probabilities $\eta \in (0, 1]$ (test sensitivity) and $\nu \in (0, 1]$ (test specificity), respectively, and independent of all other variables.
\end{asm}

After replacing the perfect-testing assumption (Simplifying Assumption~\ref{asm:perfect-test}) with the random sensitivity and specificity assumption (Assumption~\ref{asm:simple-sens-spec}), we turn to relaxing no clearance (Simplifying Assumption~\ref{asm:no-clearance}), which previously required that individuals who enter the Removed state $\set{R}$ remain there permanently.  
In real-world settings, however, individuals may re-enter the testing pool after clearance, such as through recovery, release from isolation, or return after administrative removal.

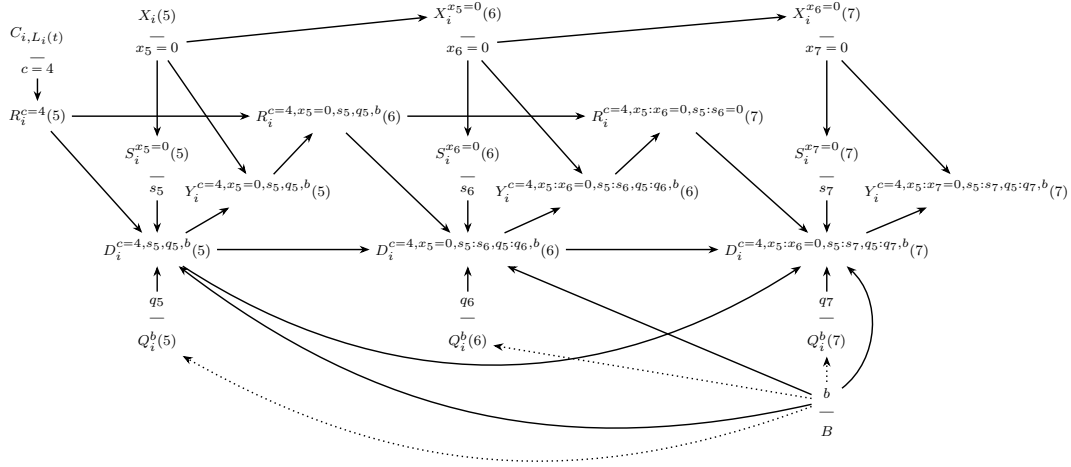
\begin{figure}[h!]
    \centering
    \resizebox{1\textwidth}{!}{
    \begin{tikzpicture}[
        node distance=0.6cm and 0.8cm,
        every node/.style={font=\scriptsize}, 
        condition/.style={draw, thick, font=\scriptsize, align=center}, 
        unobserved/.style={draw, thick, font=\scriptsize, align=center, dashed, rounded corners}, 
        dottedbox/.style={draw, thick, dotted, rounded corners},
        edge/.style={->, thick, >={Stealth[length=2mm]}}
    ]
        \node (Ri0) at (0,0) {$R_i^{c=4}(5)$};
        \node[above=1cm of Ri0] (C5) {$C_{i,L_i(t)}$}; 
        \node[below=0 of C5] (c5) {$\genfrac{}{}{0pt}{}{\rotatebox{90}{$\mid$}}{\mathlarger{c=4}}$};
        \node[below right=2cm and 0.5cm of Ri0] (Di0) {$D_i^{c=4, s_5, q_5, b}(5)$};
        \node[above=1.25cm of Di0] (Si0) {$S_i^{x_5=0}(5)$};
        \node[below=0 of Si0] (si0) {$\genfrac{}{}{0pt}{}{\rotatebox{90}{$\mid$}}{\mathlarger{s_5}}$};
        \node[above right=0.5cm and 0.0cm of Di0, xshift=-0.75cm] (Yi0) {$Y_i^{c=4, x_5=0, s_5, q_5, b}(5)$};

        \node[right=3.5cm of Ri0] (Ri1) {$R_i^{c=4, x_5=0, s_5, q_5, b}(6)$};
        \node[right=3cm of Di0] (Di1) {$D_i^{c=4, x_5=0, s_5:s_6, q_5:q_6, b}(6)$};
        \node[above=1.25cm of Di1] (Si1) {$S_i^{x_6=0}(6)$};
        \node[below=0 of Si1] (si1) {$\genfrac{}{}{0pt}{}{\rotatebox{90}{$\mid$}}{\mathlarger{s_6}}$};
        \node[above right=0.5cm and 0.0cm of Di1, xshift=-1.5cm] (Yi1) {$Y_i^{c=4, x_5:x_6=0, s_5:s_6, q_5:q_6, b}(6)$};

        \node[right=3.5cm of Ri1] (Ri2) {$R_i^{c=4, x_5:x_6 = 0, s_5:s_6 = 0}(7)$};
        \node[right=3cm of Di1] (Di2) {$D_i^{c=4, x_5:x_6=0, s_5:s_7, q_5:q_7, b}(7)$};
        \node[above=1.25cm of Di2] (Si2) {$S_i^{x_7=0}(7)$};
        \node[below=0 of Si2] (si2) {$\genfrac{}{}{0pt}{}{\rotatebox{90}{$\mid$}}{\mathlarger{s_7}}$};
        \node[above right=0.5cm and 0.0cm of Di2, xshift=-1.5cm] (Yi2) {$Y_i^{c=4, x_5:x_7=0, s_5:s_7, q_5:q_7, b}(7)$};
        
        \node[above=4cm of Di0] (Xi0) {$X_i(5)$};
        \node[above=4cm of Di1] (Xi1) {$X_i^{x_5 = 0}(6)$};        
        \node[above=4cm of Di2] (Xi2) {$X_i^{x_6 = 0}(7)$};

        \node[below=0 of Xi0] (xi0) {$\genfrac{}{}{0pt}{}{\rotatebox{90}{$\mid$}}{\mathlarger{x_5=0}}$};
        \node[below=0 of Xi1] (xi1) {$\genfrac{}{}{0pt}{}{\rotatebox{90}{$\mid$}}{\mathlarger{x_6=0}}$};
        \node[below=0 of Xi2] (xi2) {$\genfrac{}{}{0pt}{}{\rotatebox{90}{$\mid$}}{\mathlarger{x_7=0}}$};

        \node[below=0.5cm of Di0] (qi0) {$\genfrac{}{}{0pt}{}{\mathlarger{q_5}}{\rotatebox{270}{$\mid$}}$};
        \node[below=0 of qi0] (Qi0) {$Q_i^{b}(5)$};

        \node[below=0.5cm of Di1] (qi1) {$\genfrac{}{}{0pt}{}{\mathlarger{q_6}}{\rotatebox{270}{$\mid$}}$};
        \node[below=0 of qi1] (Qi1) {$Q_i^{b}(6)$};

        \node[below=0.5cm of Di2] (qi2) {$\genfrac{}{}{0pt}{}{\mathlarger{q_7}}{\rotatebox{270}{$\mid$}}$};
        \node[below=0 of qi2] (Qi2) {$Q_i^{b}(7)$};
        
        \node[below=0.5cm of Qi2] (b) {$\genfrac{}{}{0pt}{}{\mathlarger{b}}{\rotatebox{270}{$\mid$}}$};
        \node[below=0cm of b] (B) {$B$};

        \draw[edge] (Ri0) -- (Di0);
        \draw[edge] (Ri0) -- (Ri1);
        \draw[edge] (Ri1) -- (Ri2);
        \draw[edge] (Di0) -- (Yi0);
        \draw[edge] (Yi0) -- (Ri1);
        \draw[edge] (Yi1) -- (Ri2);
        \draw[edge] (Ri1) -- (Di1);
        \draw[edge] (Di1) -- (Yi1);
        \draw[edge] (Di0) -- (Di1);
        \draw[edge] (Di1) -- (Di2);
        \draw[edge] (Ri2) -- (Di2);
        \draw[edge] (Di2) -- (Yi2);
        \draw[edge] (xi0)  -- (Yi0);
        \draw[edge] (xi1) -- (Yi1);
        \draw[edge] (xi2) -- (Yi2);
        \draw[edge] (xi0)  -- (Si0);
        \draw[edge] (xi1) -- (Si1);
        \draw[edge] (xi2) -- (Si2);
        \draw[edge] (xi0) -- (Xi1);
        \draw[edge] (xi1) -- (Xi2);
        \draw[edge] (si0) -- (Di0);
        \draw[edge] (si1) -- (Di1);
        \draw[edge] (si2) -- (Di2);
        \draw[edge] (c5) -- (Ri0);    
        
        \draw[edge] (qi0) -- (Di0);
        \draw[edge] (qi1) -- (Di1);
        \draw[edge] (qi2) -- (Di2);
        \draw[edge, bend right=33] (Di0) to (Di2);
        \draw[edge, bend left=25] (b) to (Di0);  
        \draw[edge] (b) to (Di1);
        \draw[edge, bend right=50] (b) to (Di2);        
        \draw[edge, dotted, bend left=25] (b) to (Qi0);  
        \draw[edge, dotted] (b) to (Qi1);
        \draw[edge, dotted] (b) to (Qi2);
    \end{tikzpicture}
    }
    \caption{
    Single World Intervention Graph (SWIG) at $t = 5, 6, 7$, under interventions on exposure history $x_{5:t}$, symptom history $s_{5:t}$,  contact tracing history $q_{5:t}$, and shared circumstance $b$, with clearance time fixed at $C_{i,L_i(t)} = 4$. 
    Dotted lines indicate that the relationship between two variables is indirect and operates through one or more intermediate variables, rather than representing a direct causal link.
    The intervention $C_{i,L_i(t)} = 4$ fixes the most recent clearance to occur at time 4, blocking all causal pathways from prior exposures and infections to future testing at $t = 5, 6, 7$. }
    \label{fig:SWIG.clear.0}
\end{figure}

We define $C_{i,\ell}$ as the $\ell$-th time individual $i$ is cleared to re-enter the monitored population at the next timepoint. In other words, if individual $i$ is in the Removed compartment at time $C_{i,\ell}$ (i.e., $i \in R(C_{i,\ell})$), then they return to the Well compartment at time $C_{i,\ell} + 1$ (i.e., $i \in W(C_{i,\ell} + 1)$).
For any time $t$, let $L_i(t)$ denote the index of the most recent clearance before $t$, defined as $L_i(t) := \max \{ \ell : C_{i,\ell} < t \}$.
The quantity $C_{i,L_i(t)}$ denotes the most recent clearance time before time $t$ for individual $i$. 
Suppose $C_{i,L_i(t)} = 4$. 
Figure~\ref{fig:SWIG.clear.0} presents the Single World Intervention Graph (SWIG) over time points $t = 5, 6, 7$, under interventions that fix the exposure history $x_{5:7}$, symptom history $s_{5:7}$, contact tracing history $q_{5:7}$, and shared circumstance $B = b$, with the clearance time set to $C_{i, L_i(t)} = 4$.
In infectious disease settings, exposures $X_i(t)$ may arise through transmission from other individuals' prior infectiousness; by intervening on $x_{5:7}$, we block transmission induced causal pathways into variables at times $t=5,6,7$ for individual $i$ that operate through $X_i(5:7)$.
This figure also illustrates how intervening on the most recent clearance time at $C_{i,L_i(t)}=4$ isolates the causal pathways linking testing, symptoms, contact tracing, and exposures over $t=5,6,7$.

\subsection{Horvitz-Thompson Estimator}
We introduce the Horvitz–Thompson (HT) estimator for $W_+(t)$, which adjusts for individual testing probabilities that may depend on symptom history, contact tracing, clearance status, and shared circumstances.
For any time $t$ and clearance time $c$, define the full post-clearance history as
\[
\vec{\Sigma}(c,t)
= \bigl(\vec{S}(c+1:t),\, \vec{Q}(c+1:t),\, B,\, \vec{C}_{\vec{L}(t)}\bigr),
\]
and for individual $i$,
\[
\Sigma_i(c,t)
= \bigl(S_i(c+1:t),\, Q_i(c+1:t),\, B,\, C_{i,L_i(t)}\bigr).
\]
We denote the realizations of these random vectors by
\[
\vec{\sigma}_{c,t}
= \bigl(\vec{s}_{c+1:t},\, \vec{q}_{c+1:t},\, b,\, \vec{c}\bigr),
\qquad
\sigma_{i,c,t}
= \bigl(s_{c+1:t},\, q_{c+1:t},\, b,\, c\bigr),
\]
where $\vec{s}_{c+1:t}, \vec{q}_{c+1:t} \in \{0,1\}^{m \times (t-c)}$ represent the symptom and contact-tracing histories for all $m$ individuals from time $c+1$ through $t$, and $s_{c+1:t}, q_{c+1:t} \in \{0,1\}^{t-c}$ represent the corresponding histories for individual $i$.

\begin{asm}[Independence of Testing from Others' States (ITOS)]
  \label{asm:itos}
  \begin{equation*}
  \begin{aligned}
    \P\left[D_{i}(t) = 1 \mgiv \vec{W}(t), \vec{\Sigma}(c,t) = \vec{\sigma}_{c,t} \right] = \P\left[D_i(t) = 1 \mid W_i(t) = 1, \Sigma_i(c,t) = \sigma_{i,c,t}\right].
  \end{aligned}
  \end{equation*}
\end{asm}

\begin{asm}[Positivity]
  \label{asm:positivity}
  For any time $t$ and clearance time $c$, and for any realization 
  $\sigma_{i,c,t} = (s_{c+1:t}, q_{c+1:t}, b, c)$ of the stratum variable 
  $\Sigma_i(c,t)$ with $s_{c+1:t}, q_{c+1:t} \in \{0,1\}^{t-c}$,
  \[
  \P\left[D_i(t) = 1 \mid W_i(t) = 1, \Sigma_i(c,t) = \sigma_{i,c,t} \right] > 0.
  \]
\end{asm}

Assumption~\ref{asm:itos} generalizes Assumption~\ref{asm:s-itos}, and Assumption~\ref{asm:positivity} generalizes Assumption~\ref{asm:s-positivity}, by incorporating full symptom and contact tracing histories as well as clearance and shared circumstance into the testing mechanism.
As with Assumption~\ref{asm:s-positivity}, violations imply that certain subgroups are never tested, potentially biasing prevalence estimates. 
This assumption also requires modifying how scheduled testing interacts with symptom- and contact-based testing: for instance, if either symptomatic status or contact tracing guarantees testing, then an otherwise “once-per-week” schedule must still allow for a scheduled test even if a symptom- or contact-based test has already occurred earlier in the same week.

\begin{thm}[Unbiased Estimator of Prevalence]
  \label{thm:unbiased-estimator}
  Assume non-perfect testing (Assumption~\ref{asm:simple-sens-spec}), independence of testing from others' states (ITOS, Assumption~\ref{asm:itos}), and positivity (Assumption~\ref{asm:positivity}).  
  Let $\omega_i(t) = 1/\P\left[D_i(t) = 1 \mid W_i(t) = 1, \Sigma_i(c,t) = \sigma_{i,c,t} \right]$. Then,
\small{
\begin{equation}
\label{eq:adjusted-HT-estimator}
    \begin{aligned}
    & W_+(t) = \E\Bigg[ \frac{1}{\eta + \nu - 1} \sum_i \omega_i(t) D_i(t) \{1 - Y_i(t)\} - \frac{1-\eta}{\eta + \nu - 1} \sum_i \omega_i(t) D_i(t) \Big| \vec{W}(t), \vec{\Sigma}(c,t) = \vec{\sigma}_{c,t} \Bigg].
    \end{aligned}
\end{equation}
}
\end{thm}

Theorem~\ref{thm:unbiased-estimator} provides an unbiased Horvitz–Thompson estimator for $W_+(t)$ under the full timeline in Figure~\ref{fig:flowchart}. 
The proof is provided in the Appendix and follows the same structure as that of Theorem~\ref{thm:s-unbiased-estimator}.

\subsection{Computing Testing Probabilities from a Counterfactual Scenario}
Accurate estimation of the testing probability in Theorem~\ref{thm:unbiased-estimator} is crucial for unbiased inference under the Horvitz–Thompson framework. 
Although $\Sigma_i(c,t)$ are observed, this probability is not directly identifiable because $W_i(t)$ is only partially observed.  
To address this, we define a counterfactual testing probability (Theorem~\ref{thm:testing probability}).

\begin{thm}[Estimation of Testing Probability]
  \label{thm:testing probability}
  Assume non-perfect testing (Assumption~\ref{asm:simple-sens-spec}), consistency (Assumption~\ref{asm:s-consist}), and exchangeability (Assumption~\ref{asm:s-exch}). Then
  \small{
  \begin{equation}
    \label{Aeq:identifiable-testing-prob}
    \begin{aligned}
      & \P\left[D_i(t) = 1 \mid W_i(t) = 1, \Sigma_i(c,t) = \sigma_{i,c,t} \right] \\
      &=\frac{
        \P[D_i^{x_{c+1:t-1} = 0, \sigma_{i,c,t}}(t) = 1, R_i^{x_{c+1:t-1} = 0, \sigma_{i,c,t}}(t) = 0]
      }{        
      \P[R_i^{x_{c+1:t-1} = 0, \sigma_{i,c,t}}(t) = 0]
      } 
    \end{aligned}
   \end{equation}
   }
\end{thm}

Theorem~\ref{thm:testing probability} shows that the probability of testing at time $t$, conditional on being Well and given symptom history, contact tracing history, clearance status, and shared circumstance, can be expressed as a ratio of counterfactual probabilities under an intervention fixing $x_{c+1:t-1}$ and $\sigma_{i,c,t}$.
This identifies the testing probability despite $W_i(t)$ being unobserved in real data.
If the testing mechanism is fully known (e.g., governed by a specified protocol), this counterfactual probability can be computed directly.  
Otherwise, it must be estimated from observed data, as discussed in the following section.

\subsection{Estimating Counterfactual Testing Probabilities from Data}
If the testing mechanism is not fully specified (e.g., depends on unobserved rules or shared circumstance), the counterfactual probability in Equation~\eqref{Aeq:identifiable-testing-prob} cannot be derived from design and must instead be estimated from observed data under causal assumptions.
The numerator of Equation~\eqref{Aeq:identifiable-testing-prob} decomposes as:
\begin{equation}
\begin{aligned}
    &\P\Big[D_i^{x_{c+1:t-1} = 0, \sigma_{i,c,t}}(t) = 1, R_i^{x_{c+1:t-1} = 0, \sigma_{i,c,t}}(t) = 0 \Big] \\
    &= \sum_{u=c}^{t-1} 
    \Big( 
        \P\Big[Z_{i, K_i(u+1)+1}^{x_{c+1:t-1} = 0, \sigma_{i,c,t}} = t, R_i^{x_{c+1:t-1} = 0, \sigma_{i,c,t}}(t) = 0 
        \mid D_i^{x_{c+1:u-1} = 0, \sigma_{i,c,u}}(u) = 1\Big]  \\
    &\hspace{1.5cm}  \times \P\Big[ D_i^{x_{c+1:u-1} = 0, \sigma_{i,c,u}}(u) = 1 \Big] 
    \Big),
\end{aligned}
\end{equation}
where the first term in the summation becomes:
\begin{equation}
\begin{aligned}
\label{Aeq:counter-testing-prob}
    &\P \left[
        Z_{i,K_i(u+1)+1}^{x_{c+1:t-1}=0, \sigma_{i,c,t}} = t, 
        R_i^{x_{c+1:t-1}=0, \sigma_{i,c,t}}(t) = 0 \right. \\
    &\hspace{2cm} \left. \mid D_i^{x_{c+1:u-1}=0, \sigma_{i,c,u}}(u) = 1, 
        Y_i^{x_{c+1:u}=0, \sigma_{i,c,u}}(u) = 0 
    \right] \times \nu,
\end{aligned}
\end{equation}
where $\nu = \P \left[ Y_i^{x_{c+1:u}=0, \sigma_{i,c,u}}(u) = 0  \mid D_i^{x_{c+1:u-1}=0, \sigma_{i,c,u}}(u) = 1 \right]. $
Then, omitting $\nu$ for simplicity, the conditional probability in Equation~\eqref{Aeq:counter-testing-prob} can be equivalently written as
\begin{equation}
   h^{x_{c+1:t-1}=0, \sigma_{i,c,t}}(t) \times \prod_{\tau=u+1}^{t-1} 
\left( 1 -  h^{x_{c+1:\tau-1}=0, \sigma_{i,c,\tau}}(\tau) \right), 
\end{equation}
where the counterfactual conditional hazard is defined in Theorem~\ref{thm:hazard-link}.

\begin{thm}[Equivalence of Counterfactual and Factual Conditional Hazard Functions with Symptom and Contact Tracing]
\label{thm:hazard-link}
Under consistency (Assumption~\ref{asm:s-consist}) and exchangeability (Assumption~\ref{asm:s-exch}),  
the counterfactual conditional hazard function  under no prior exposure and fixed symptom history, tracing history, clearance, and shared circumstance,  
$ h^{x_{c+1:\tau-1}=0, \sigma_{i,c,\tau}}(\tau)$, is equal to the corresponding factual conditional hazard function $h^{\text{obs}}(\tau)$, for any $u \in \{c+1, \ldots, t-1\}$ and $\tau \in \{u+1, \ldots, t\}$:
\small{
\begin{equation}
  h^{x_{c+1:\tau-1}=0, \sigma_{i,c,\tau}}(\tau) = h^{\text{obs}}(\tau),
\end{equation}
}
where $ h^{x_{c+1:\tau-1}=0, \sigma_{i,c,\tau}}(\tau)$ is defined as
\small{
\begin{equation}
\begin{aligned}
 &\P\left[
        Z_{i,K_i(u+1)+1}^{x_{c+1:\tau-1}=0, \sigma_{i,c,\tau}} = \tau,
        R_i^{x_{c+1:\tau-1}=0, \sigma_{i,c,\tau}}(\tau) = 0 \mid \right. \\
    &\hspace{1cm} \left.    
        Z_{i,K_i(u+1)+1}^{x_{c+1:\tau-1}=0, \sigma_{i,c,\tau}} \geq \tau, 
        D_i^{x_{c+1:u-1}=0, \sigma_{i,c,u}}(u) = 1,
        Y_i^{x_{c+1:u}=0, \sigma_{i,c,u}}(u) = 0
    \right] \\
\end{aligned}
\end{equation}
}
and $h^{\text{obs}}(\tau)$ is defined as 
\small{
\begin{equation}
\begin{aligned}
 &\P\Big[ Z_{i,K_i(u+1)+1} = \tau, R_i(\tau) = 0 \mid Z_{i,K_i(u+1)+1} \geq \tau, D_i(u) = 1, Y_i(u) = 0,
 \Sigma_i(c,\tau) = \sigma_{i,c,\tau}  \Big]. 
\end{aligned}
\end{equation}
}
\end{thm}

\begin{corollary}
\label{cor:test-probs-sympt}
Assume non-perfect testing (Assumption~\ref{asm:simple-sens-spec}), consistency (Assumption~\ref{asm:s-consist}), and exchangeability (Assumption~\ref{asm:s-exch}).  
The numerator and denominator of the counterfactual testing probability under the intervention
$x_{c+1:t-1}=0, \sigma_{i,c,t}$
are expressed as a ratio of two factual probabilities, with the numerator and denominator defined respectively as follows:
\small{
\begin{equation}
\begin{aligned}
    &\P\Big[D_i^{x_{c+1:t-1} = 0, \sigma_{i,c,t}}(t) = 1, R_i^{x_{c+1:t-1} = 0, \sigma_{i,c,t}}(t) = 0 \Big] \\
    &= \sum_{u=c}^{t-1} 
    \Big( 
        \P\Big[
        Z_{i,K_i(u+1)+1} = t, R_i(t) = 0 \mid D_i(u) = 1, Y_i(u) = 0, 
          \Sigma_i(c,t) = \sigma_{i,c,t}  \Big] \times \nu  \\
    &\hspace{1cm} \times \P\Big[ D_i^{x_{c+1:u-1} = 0, \sigma_{i,c,u}}(u) = 1 \Big] 
    \Big),
\end{aligned}
\end{equation}
}
and 
\small{
\begin{equation}
\begin{aligned}
    &\P\Big[R_i^{x_{c+1:t-1} = 0, \sigma_{i,c,t}}(t) = 0 \Big] \\
    &= \sum_{u=c}^{t-1} 
    \Big( 
        \P\Big[
        Z_{i,K_i(u+1)+1} \geq t, R_i(t) = 0 \mid D_i(u) = 1, Y_i(u) = 0, 
         \Sigma_i(c,t) = \sigma_{i,c,t}  \Big] \times \nu  \\
    &\hspace{1cm} \times \P\Big[ D_i^{x_{c+1:u-1} = 0, \sigma_{i,c,u}}(u) = 1 \Big] \Big).
\end{aligned}
\end{equation}
}
Under the stated assumptions, each term on the right-hand side is identifiable from the observed data. 
\end{corollary}

Theorem~\ref{thm:hazard-link} and Corollary~\ref{cor:test-probs-sympt} are proved in the Appendix.
Theorem~\ref{thm:hazard-link} employs consistency and exchangeability to replace counterfactual terms with their observed counterparts, and Corollary~\ref{cor:test-probs-sympt}, which follows directly from the theorem, establishes that the counterfactual testing probability in Equation~\eqref{Aeq:identifiable-testing-prob} is identifiable from observed data.

The preceding identification result has three implications for implementing the HT estimator.
First, when the counterfactual testing probabilities in Equation~(15) are known, the corresponding HT estimator is unbiased because the inverse probability weights are computed from the true testing probabilities.
This remains true even when some realized strata are untested on a given day, because an untested realized stratum does not necessarily imply that the underlying testing probability in that stratum is zero.
Thus, the positivity assumption is a prospective condition on the underlying testing probability in each stratum before testing is realized, not a condition on the realized proportion tested in the observed sample.
This distinction is supported by \citet{schnell2024overcoming}, who showed, in a simpler setting, that an HT estimator with correctly specified testing probabilities is unbiased even when some realized strata are untested.
Second, in practice, especially when testing includes symptomatic testing and contact tracing in addition to scheduled testing, the counterfactual testing probabilities in Equation~(15) are generally difficult to know exactly and must be estimated from the observed data.
In this setting, bias arises from Jensen's inequality because the inverse of an unbiased estimator of testing probability is not an unbiased estimator of the inverse of the true testing probability.
As the population size increases while the tested proportion is held fixed, the estimated testing probabilities converge to their true values by the law of large numbers, and this Jensen bias vanishes asymptotically.
Third, for realized strata with no tested individuals, the within-stratum prevalence is not nonparametrically estimable from the observed data, so an implementation rule is required.
For the primary implementation, we assume that nonremoved individuals in such untested strata are Well. 
This assumption is reasonable in our setting because the overall prevalence is low, and these untested strata tend to consist of individuals who were tested more recently than individuals in other strata, suggesting that their prevalence may be lower than the overall prevalence.
We also conducted additional sensitivity analyses by considering alternative assumptions that half or none of the nonremoved individuals in the untested stratum were Well, to assess whether the empirical conclusions are sensitive to this implementation choice.

In this section, we extend the core framework to incorporate symptom status, contact tracing eligibility, clearance, and shared circumstance into the counterfactual outcomes formulation and the corresponding Directed Acyclic Graph (DAG). 
Symptom status and contact tracing eligibility are intermediate variables influenced by latent infection or exposure status and, in turn, affect the probability of being tested. 
Shared circumstance captures contextual factors, such as group-level behaviors or institutional policies, that jointly influence multiple individuals’ testing. 
These additions create new dependencies between exposure and testing, which are explicitly represented in the DAG to reveal potential bias pathways. 
By formalizing these relationships, the extended framework specifies the conditions under which prevalence can be identified and estimated without bias, even in the presence of complex testing mechanisms.

\section{Simulation Studies}
\label{sec:simulation}

\subsection{General Setup}
We conducted a simulation study to evaluate the performance of three estimators across four scenarios with symptomatic and contact tracing representation, extending the framework of \citet{schnell2024overcoming}.
The simulation design was inspired by longitudinal testing data from 11,335 undergraduate students residing on campus at The Ohio State University during the Fall 2020 semester. 
The general simulation parameters were as follows.

A population of $10,000$ individuals with identically distributed processes was simulated over a 21-day period. 
Individuals were organized into $5000$ exchangeable clusters, with 2 exchangeable individuals per cluster, representing living situations with roommates.
The hazard of initial exposure from outside the cluster for individuals in the nonremoved state was modeled as 
$h(\tau) = \frac{1}{10} \left( \frac{\tau (21 - \tau)}{(21 / 2)^2} \left( \frac{1}{10} - \frac{1}{50} \right) + \frac{1}{50} \right),$
where $\tau$ denotes the time since day 0 or since the most recent clearance. 
The within-cluster exposure hazard was set to $1/5$ multiplied by the number of infectious individuals in the same cluster, and was assumed to act independently of external exposure.
The hazard for subsequent exposures (i.e., after prior infection) was defined as $\frac{1}{2}h(\tau)$. 
Simulations were initialized with a 2\% prevalence and typically peaked at around 5\%.
Test sensitivity and specificity were set to 83.2\% and 99.2\%, respectively, based on values reported in the meta-analysis of saliva-based PCR tests for SARS-CoV-2 by \citet{butler2021comparison}. 
Following a positive test, individuals entered the Removed compartment for a period of 5 days, after which they re-entered the Well compartment.

On each simulation day $t \in \{1, \ldots, 21\}$, we evaluate the Horvitz-Thompson (HT) estimator of prevalence using estimated testing probabilities from observed data.  
The true prevalence in each replicate $n \in \{1, \ldots, 100\}$ is defined as
$p^{(n)}(t) 
= \frac{I_+^{(n)}(t)}{M - R_+^{(n)}(t)}
= \frac{M - R_+^{(n)}(t) - W_+^{(n)}(t)}{M - R_+^{(n)}(t)},$
where $M = 10{,}000$ denotes the total population size. 
The unobserved quantity $W_+^{(n)}(t)$ is replaced by its Horvitz-Thompson estimate, as established in Theorem~\ref{thm:unbiased-estimator} and Corollary~\ref{cor:test-probs-sympt}, yielding the estimated prevalence computed from observed data using the estimated testing weights.
In our implementation, we estimate these weights using joint stratification on $(c,s)$, where $c$ is the last clearance time and $s$ is the last symptomatic or contact-tracing test time. 
This stratification is used because $(c,s)$ captures the key components of an individual's recent testing history that most directly influence their probability of being tested.
We construct 95\% confidence intervals using the delete a group (grouped) jackknife of \citet{kott2001delete}, which is computationally less demanding than the standard (leave one unit out) jackknife of \citet{tukey1958bias} and than the bias corrected and accelerated bootstrap of \citet{efron1987better}.

\subsection{Testing Scenario}

We first describe scheduled testing with pre-defined rules, followed by symptom-based and contact tracing–based testing triggered by symptoms or confirmed exposures.

\begin{itemize}
  \item \textit{Simple random testing regimen:} nonremoved individuals are tested independently each day with probability $1/6$.

  \item \textit{Once-per-period regimen:} each nonremoved individual is eligible for one scheduled test per fixed-length calendar interval (e.g., every 7 days). The testing time is drawn uniformly within the interval, with the daily testing probability increasing as the interval progresses to ensure one scheduled test per period.

  \item \textit{Max-gap regimen:} the time of first scheduled test is uniformly distributed over the first 10 days. On subsequent days $t$, if the most recent scheduled test or clearance time is $z$, nonremoved individuals are tested with probability $(t - z)^2 / 10^2$.

  \item \textit{Min-max regimen:} similar to the max-gap regimen, but testing is prohibited within 5 days of the most recent scheduled test.
\end{itemize}

In the simulation, individuals exposed to COVID-19 become infectious and develop symptoms on their first infectious day with probability $0.25$, leading to immediate testing regardless of standard eligibility rules. 
Unexposed individuals may also report symptoms with probability $0.01$, reflecting non-COVID-19 conditions such as seasonal illness or allergies, which likewise trigger testing under the symptom-based mechanism. 
This extends the framework of \citet{schnell2024overcoming}, which considered only COVID-19 symptoms and did not allow for symptom-driven testing among uninfected individuals. 
In the contact tracing scenario, whenever an individual tests positive, all other nonremoved members of their cluster are tested the following day, again independent of standard eligibility rules.

\begin{figure}[h!]
    \centering
    \includegraphics[width=1\textwidth]{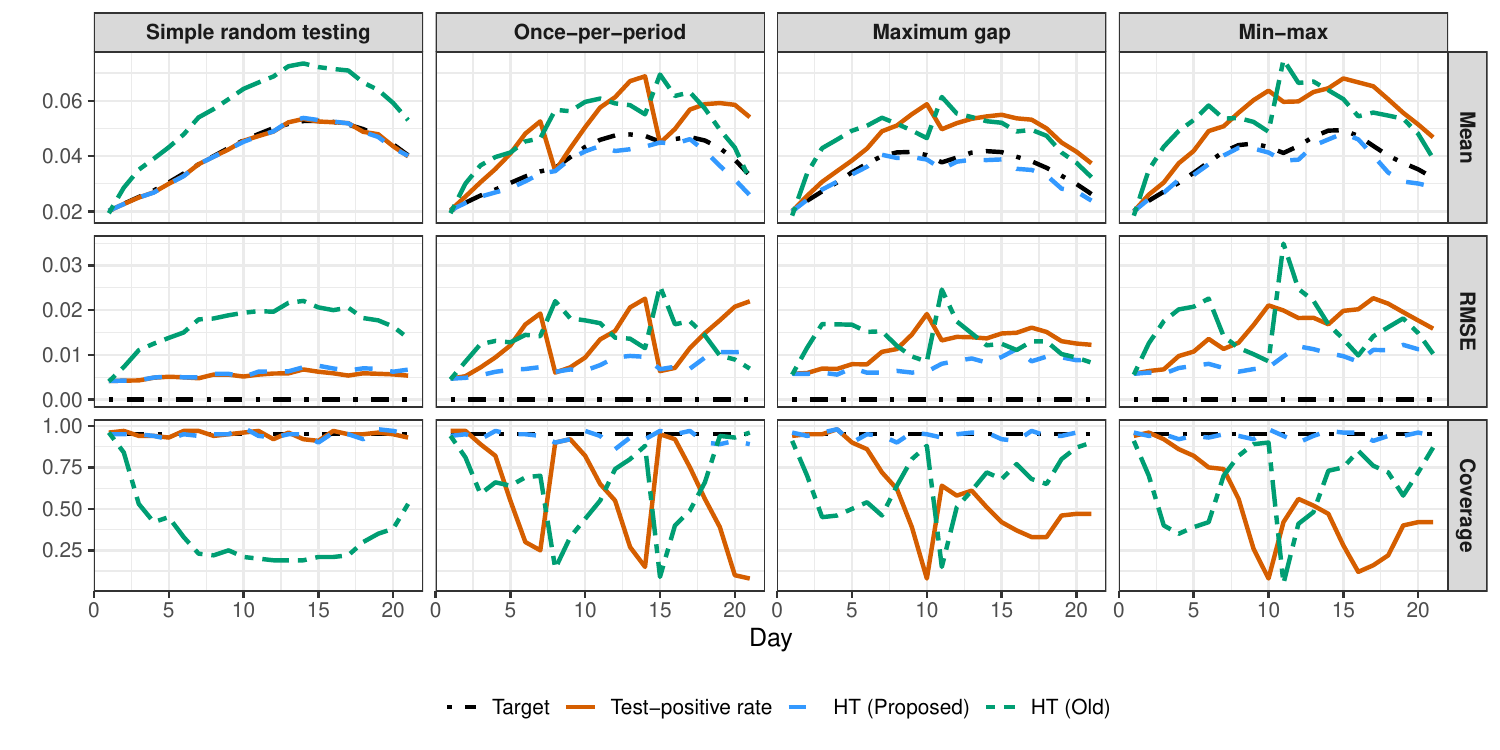}
    \caption{Prevalence estimation under four scheduled testing designs with additional symptomatic and contact-tracing testing. 
    For each design (columns), panels display the mean (top), RMSE (middle), and coverage probability (bottom) over 100 simulation replicates. 
    Curves correspond to the target prevalence (black dotdash), the test-positive rate (red solid), the HT (Proposed) estimator (blue dashed), and the HT (Old) estimator (green twodash). 
    Coverage probabilities are computed from 95\% confidence intervals constructed using a delete-a-group (grouped) jackknife with 20 groups for all estimators.
    }
    \label{fig:sim-prev}
\end{figure}

Figure~\ref{fig:sim-prev} presents simulation results for prevalence estimation under testing designs that include both symptomatic and contact tracing tests. 
The figure compares three estimators: the test-positive rate (red solid), the HT (Proposed) estimator (blue dashed), and the naive HT (Old) estimator of \citet{schnell2024overcoming} (green twodash), which treats all tests as scheduled; the target prevalence is shown in black dotdash.
Although the test-positive rate accounts for imperfect testing and symptom or contact tracing history (see Appendix for derivation), it remains substantially biased relative to the target prevalence and yields consistently higher RMSE across once per period, maximum gap, and min-max strategies, generally overestimating the true prevalence. 
In contrast, the HT (Proposed) estimator is nearly unbiased with much lower RMSE. 
The naive HT (Old) estimator shows the opposite behavior, with pronounced upward bias under all designs and a substantial tendency to overestimate the true prevalence.
Moreover, although not shown here, when we apply the HT (Old) estimator after removing contact tracing tests and symptomatic tests (rather than treating them as scheduled tests), the estimator displays a very large downward bias, which is even more severe than the bias obtained when all tests are treated as scheduled.
Across all four scenarios, the HT (Proposed) estimator attains coverage close to the nominal 95\% level, while the other estimators show pronounced undercoverage, except for the test-positive rate under simple random testing.
Both the naive HT (Old) estimator and the HT (Proposed) estimator assume that nonremoved individuals in realized untested strata are Well.
Additional strata diagnostics in the Appendix show that such untested strata are virtually absent early in follow-up and arise mainly later; when they appear, they are typically very small, often containing fewer than 5 nonremoved individuals, with a total size of about 100 or fewer at each time point considered.
Sensitivity analyses under alternative assumptions for these untested strata yield similar results when all or half of their members are assumed to be Well, whereas assuming that none are Well leads to less favorable performance, although it still outperforms the HT (Old) estimator and the test-positive rate.

In practice, when factors affecting both testing and exposure are present, the HT (Proposed) estimator can be implemented with additional conditioning or stratification variables to account for such heterogeneity. 
For example, residence-related factors, such as membership in a certain dormitory, as well as year in school, may influence both testing behavior and infection risk and should therefore be incorporated when relevant. 
Motivated by this concern, we conducted an additional simulation study in the Appendix with two heterogeneous dormitory subpopulations that differed in exposure risk and testing behavior. 
In this setting, one dormitory had lower exposure risk and less frequent scheduled testing, with no symptom-based testing or contact tracing, whereas the other followed the original testing design with symptom-based testing and contact tracing. 
We compared a pooled test-positive rate and a pooled HT (Proposed) estimator, each applied to the full population, with a stratified HT (Proposed) estimator defined as a weighted average of dormitory-specific HT estimates, with time-specific weights given by each dormitory's proportion among all nonremoved individuals. 
Across the four testing designs, the stratified estimator remained approximately unbiased and often had smaller RMSE with coverage closer to the nominal level, whereas the pooled HT estimator showed greater bias and variability in several settings and the pooled test-positive rate exhibited even larger bias than both HT estimators. 
This analysis directly examines performance when the identically distributed assumption is violated by heterogeneous subpopulations with different testing mechanisms.
We also examined robustness to misspecification of the sensitivity parameter in the Appendix.
Because specificity is typically very high for COVID-19 testing, we focused on sensitivity as the more practically consequential source of misspecification. 
As expected, using a lower assumed sensitivity generally increased the estimated prevalence, whereas using a higher assumed sensitivity generally decreased it, with larger changes occurring as the assumed value moved farther from the correctly specified value. 
Misspecification of the sensitivity parameter changes all three estimators by a broadly similar amount, but the HT (Proposed) estimator is the only estimator that is approximately unbiased when the sensitivity parameter is correctly specified.
Together, these results highlight the importance of accounting for major sources of population heterogeneity and carefully specifying test accuracy parameters in applied settings.

\section{Prevalence Estimates from OSU Fall 2020}
\label{sec:realdata}

We analyzed de-identified longitudinal testing data from 11,335 undergraduate students residing on campus during the Fall 2020 semester. 
Testing was carried out by three providers: Student Health Services (SHS), Vault Health (VAULT), and the Applied Microbiology Services Laboratory (AMSL). 
All eligible students were required to complete one saliva-based PCR test per calendar workweek (Monday–Friday), selecting their testing day within that period through either VAULT or AMSL. 
SHS primarily conducted additional tests for symptomatic individuals and those identified through contact tracing. 
During the early move-in period (August 14–16) and prior to the start of classes, all students were tested upon arrival regardless of symptoms, and tests administered by SHS and VAULT were not distinguished.
On average, symptomatic or contact-tracing tests accounted for approximately 1-2\% of all daily tests, with the total number of daily tests typically ranging from 1,000 to 2,000.
We assume a clearance period of 10 days following a positive test, during which individuals were removed from testing, followed by an additional 80-day exemption from scheduled surveillance. 
Test sensitivity and specificity were set to 0.832 and 1, respectively. 
The prevalence lower bound was truncated at zero.

\begin{figure}[h!]
    \centering
    \includegraphics[width=1\textwidth]{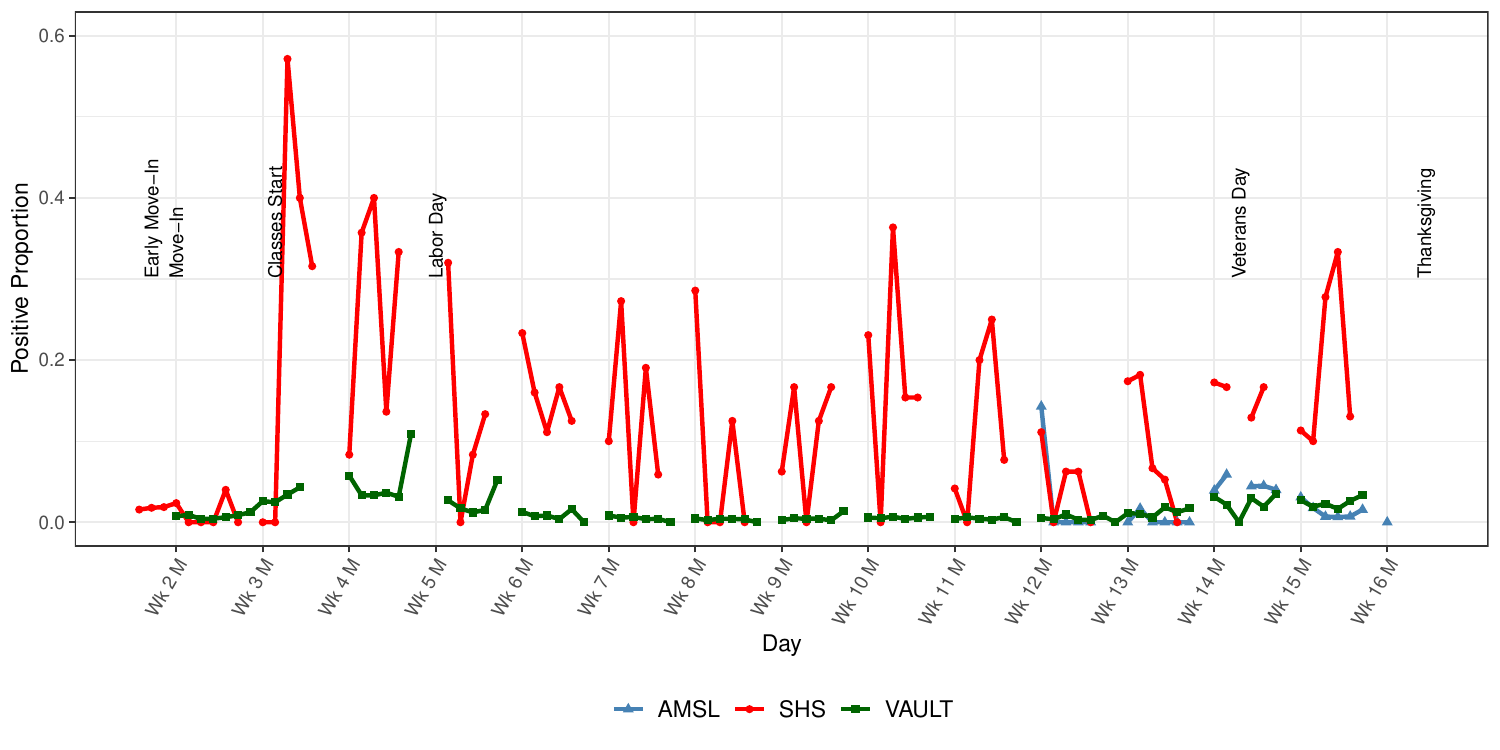}
    \caption{Daily naïve test-positive rates by provider.  
    SHS primarily corresponds to symptomatic and contact-tracing testing, whereas AMSL and VAULT correspond to scheduled surveillance testing.  
    During the early move-in period and before classes started, tests conducted by SHS and VAULT were not distinguished.  
    There was little to no testing conducted on Saturdays, Sundays, or holidays.}
    \label{fig:tpr-provider}
\end{figure}

As an exploratory analysis, Figure~\ref{fig:tpr-provider} displays the naïve test-positive rates stratified by testing provider (AMSL, SHS, and VAULT). 
SHS shows substantially higher positivity throughout the semester, consistent with its role in symptomatic and contact-tracing testing.  
By contrast, AMSL and VAULT exhibit much lower positivity reflective of scheduled surveillance.  
The sharp rise in SHS positivity around the start of classes highlights the onset of systematic differentiation between symptomatic/contact-tracing and scheduled testing.  
These patterns confirm the need to account for heterogeneity in the testing mechanism when estimating prevalence.

\begin{figure}[h!]
    \centering
    \includegraphics[width=1\textwidth]{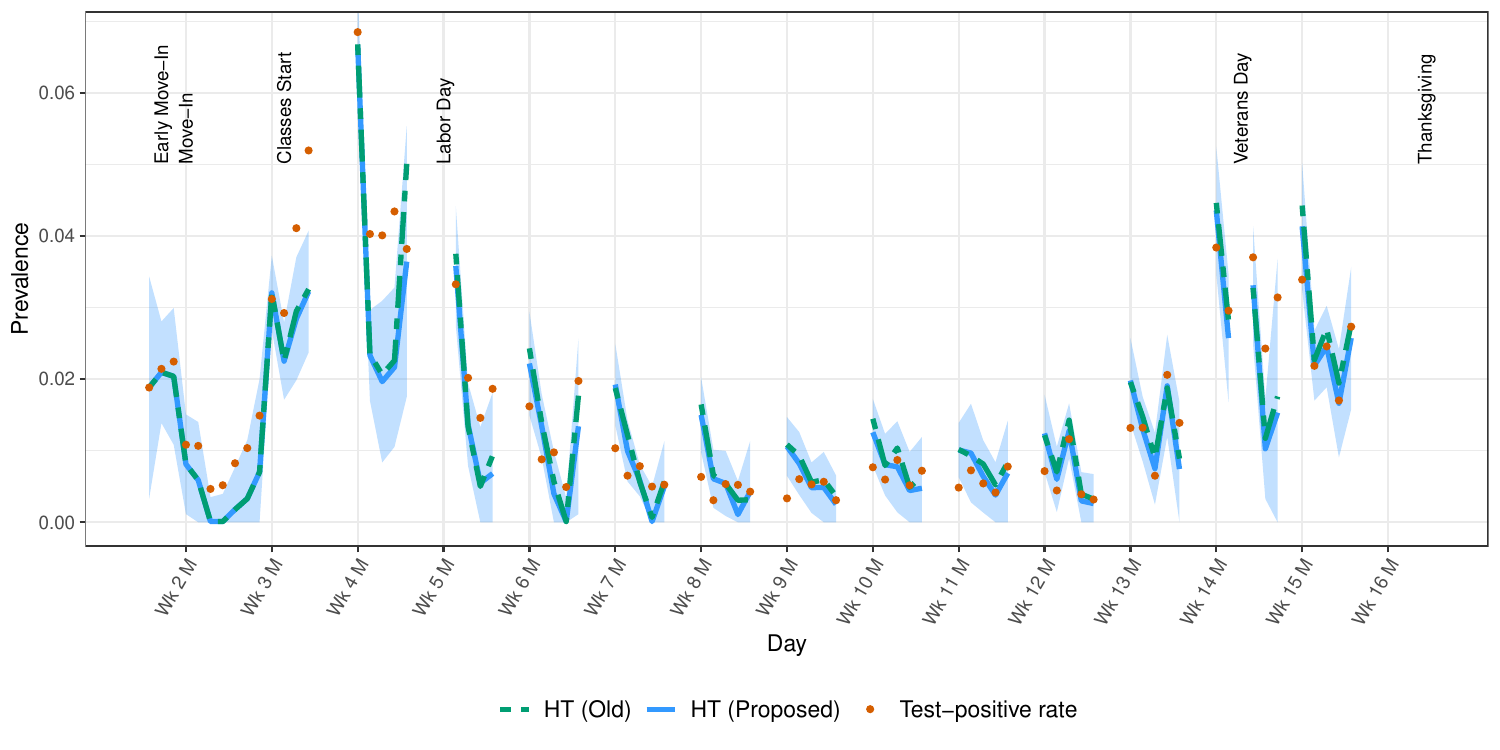}
    \caption{
    Daily prevalence estimates computed using the HT (Proposed) estimator (blue solid), the HT (Old) estimator (green twodash), and the test-positive rate (red points).
    Key events such as early move in, move in, the start of classes, Labor Day, Veterans Day, and Thanksgiving are annotated above the corresponding days.
    Vertical grid lines indicate Mondays.
    Days with fewer three hundred tests are omitted from the plot.
    Shaded ribbons show 95\% confidence intervals constructed using a delete-a-group (grouped) jackknife with 20 groups.
    }
    \label{fig:prev}
\end{figure}

Figure~\ref{fig:prev} displays the daily test-positive rates together with the Horvitz–Thompson (HT) prevalence estimates, where the HT (Proposed) estimator adjusts for symptomatic and contact tracing tests, and the HT (Old) estimator treats all tests as scheduled.
For the HT (Proposed) estimator, shaded ribbons indicate 95\% confidence intervals constructed using a delete a group jackknife with 20 groups.
Consistent with the bias patterns in Figure~\ref{fig:sim-prev}, the test-positive rate (red points) and the HT (Old) estimator (green twodash) are generally higher than the HT (Proposed) estimator (blue solid) once classes begin.
Before the start of classes, the HT (Proposed) and HT (Old) estimators coincide because different types of testing are not distinguished, that is, symptomatic and contact tracing tests are not separated from the overall testing process.

A limitation of this real-data analysis is that no external gold-standard estimate of prevalence is available for the OSU student population during the study period. 
Therefore, Figure~\ref{fig:prev} cannot by itself establish that the HT (Proposed) estimator is less biased than either the test-positive rate or the HT (Old) estimator. 
Rather, the real-data analysis is intended to illustrate how the estimators differ in practice under repeated testing with scheduled, symptomatic, and contact-tracing components. 
Based on the identification results and simulation evidence, the HT estimators are expected to be less biased than the test-positive rate because they explicitly account for the time-varying testing mechanism. 
During the middle of the semester, the test-positive rate suggests that prevalence remained relatively constant within each week, whereas the HT estimators suggest that prevalence decreased after infectious individuals were identified and removed from the testing population. 
The latter pattern is consistent with the expectation that repeated testing and removal reduced the pool of infectious students.
The remaining temporal variation may reflect infections occurring outside the testing process, including the possibility that infections occurred more often on weekends than within classroom settings. 
Although the HT (Old) and HT (Proposed) estimates are relatively close in this application, symptomatic testing and contact tracing remain natural sources of bias for the HT (Old) estimator because they induce testing that depends on infection-related histories, which the HT (Proposed) estimator is designed to address.
Thus, the claim that the HT (Proposed) estimator is less biased is supported primarily by the identification results and simulation evidence, whereas the OSU analysis provides an empirical illustration of the differences among the estimators. 
External campus or community surveillance data may provide useful qualitative context for temporal trends, but they do not provide a directly comparable prevalence benchmark for the same target population and time scale.

\section{Discussion}
\label{sec:discussion}

This work develops a unified counterfactual outcomes framework for estimating prevalence under scheduled, symptom-based, and contact-tracing testing. 
Within this framework, we show that the Horvitz-Thompson (HT) estimator, constructed using estimated conditional testing probabilities, remains approximately unbiased under both scheduled and preferential testing mechanisms. 
Simulation studies demonstrate that the HT estimator effectively adjusts for selection bias, yielding more accurate prevalence estimates than the test-positive rate.

At the same time, sparsity remains a practical limitation. 
As the number of realized strata increases over time, some strata may be untested on a given day. 
Because the proposed estimator stratifies on recency summaries rather than the full longitudinal testing history, the growth in the number of strata is less severe than under full-history stratification, although sparsity may still become an issue. 
In our simulations and OSU application, untested strata arose mainly later in follow-up and were typically small, with total size about 100 or fewer at the time points considered, suggesting limited impact in the settings considered here. 
Nevertheless, for longer surveillance periods or richer conditioning sets, some principled form of coarsening, pooling, or regularization may become useful.

Future work will extend this population-level framework to individual-level pairwise survival models \citep{kenah2011contact, kenah2015semiparametric, kenah2019pairwise, sharker2024pairwise}, which explicitly characterize infection and transmission processes between individuals. 
Such models will allow joint inference on infection hazards and testing mechanisms, providing a richer basis for evaluating surveillance strategies and intervention effects in clustered populations.

\begin{acks}[Acknowledgments]
We thank Sarah Grim for clarifying the Ohio State University COVID 19 testing process, particularly the procedures for handling contact tracing and symptomatic testing at Student Health Services.
\end{acks}

\begin{funding}
Partial funding for this research was provided by the National Science Foundation under grant DMS-2027001. 
\end{funding}

\begin{supplement}
\stitle{Supplementary Material for ``A Counterfactual Framework for Estimating Infectious Disease Prevalence under Repeated Testing with Symptomatic and Contact-Tracing Components''}
\sdescription{The supplement provides detailed proofs and additional simulation studies.}
\end{supplement}

\begin{supplement}
\stitle{Data and R Code}
\sdescription{
The human data collection and analysis was approved under The Ohio State University IRB protocol 2021H0189. 
The supplementary materials include anonymized data and R scripts to reproduce the simulation study and the applied analyses.}
\end{supplement}


\bibliographystyle{imsart-nameyear} 
\bibliography{aoas}       

\end{document}